\documentclass[11pt]{article}

\usepackage{amssymb}
\usepackage{amsmath,amsthm,amsfonts}
\usepackage{mathabx}
\usepackage{epstopdf}
\usepackage{color}
\usepackage{epsfig,amssymb,amsfonts,amsmath,graphicx,dsfont,cite,xfrac}
\usepackage[affil-it]{authblk}
\usepackage{subcaption}
\definecolor{mygray}{gray}{0.5}

\usepackage{cite}
\usepackage[colorlinks=true,linkcolor=blue,citecolor=red]{hyperref}

\title{Two-mode squeezed state quantisation and semiclassical portraits}
\author[1]{Jean-Pierre Gazeau \thanks{gazeau@apc.in2p3.fr}}
\author[2,3]{V\'eronique Hussin \thanks{hussin@dms.umontreal.ca}}
\author[3,4]{James Moran \thanks{james.moran@umontreal.ca}}
\author[5]{Kevin Zelaya \thanks{kdzelaya@fis.cinvestav.mx}}
\affil[1]{CNRS, Astroparticule et Cosmologie, Universit\'e de Paris, F-75013 Paris, France}
\affil[2]{D\'epartement de math\'ematiques et de statistique, Universit\'e de Montr\'eal, Montr\'eal,
Qu\'ebec H3C 3J7, Canada}
\affil[3]{Centre de recherches math\'ematiques, Universit\'e de Montr\'eal, Montr\'eal, Qu\'ebec H3C 3J7, Canada}
\affil[4]{D\'epartement de physique, Universit\'e de Montr\'eal, Montr\'eal, Qu\'ebec H3C 3J7, Canada}
\affil[5]{Nuclear Physics Institute, The Czech Academy of Science, \v{R}e\v{z}, 250 68 Husinec, Czech Republic}

\date{}
\begin{document}

\maketitle

\begin{abstract}
Quantisation with Gaussian type states offers certain advantages over other quantisation schemes, in particular, they can serve to regularise formally discontinuous classical functions leading to well defined quantum operators. In this work we define a squeezed state quantisation in two dimensions using several families of squeezed states for one- and two-mode configurations. The completeness relations of the squeezed states are exploited in order to tackle the quantisation and semiclassical analysis of a constrained position dependent mass model with harmonic potential. The effects of the squeezing parameters on the resulting operators and phase space functions are studied, and configuration space trajectories are compared between the classical and semiclassical models.
\end{abstract}
\newpage
\tableofcontents
\section{Introduction}
Coherent states and their generalisation, squeezed states, are ubiquitous in the study of quantum optics. They describe a set of minimal uncertainty states with respect to their generalised quadratures (in typical quantum systems these may refer to position and momentum), and the `squeezing' refers to the reduction in one quadrature variance at the expense of an increase in the conjugate quadrature variance \cite{gerryknight6}. In multimode systems there exists an even greater variety of squeezed states because the squeezing can occur between four or more quadratures and their combinations. Multimode squeezed states are the most general type of Gaussian state permissible and have found use outside of optics as a resource in continuous variable quantum information processing for generating multipartite entanglement \cite{Pfistera, addesso}. Beyond  Gaussian states, non-Gaussian states represent a further generalisation \cite{Ra2020aa}. Schumaker investigated the most general two-mode Gaussian pure states \cite{schumaker}, schemes designed to generalise this construction to the $N$-mode case were studied in \cite{xinrhodes}, and a general presentation of a coupled three-mode squeezed vacuum was presented in \cite{glasserzhang}.

Outside of optics and information theory, squeezed states have attractive mathematical properties, in particular they form an overcomplete basis in the Hilbert space of quantum states \cite{holomorphh, 2011}. Equipped with this property one may expand any state of a given system in the basis of squeezed states. Moreover, one may quantise a classical function in the squeezed state basis yielding an associated quantum operator as well as define an averaged value of the initial function with respect to the squeezed states yielding a semiclassical portrait. This is precisely the purpose of this work, to extend the ideas of Klauder-Berezin coherent state quantisation~\cite{Kla12} by defining a two-mode squeezed state quantisation in which we use several families of two-mode squeezed states and study the effect of their squeezing parameters on the resulting quantum operators and semiclassical phase space functions.

Let $A$ and $B$ be two observables,\footnote{In this work, we focus on the common definition of observables as defined by self-adjoint operators, $A^{\dagger}=A$.} so that $[A,B]\equiv AB-BA=\mathrm{i}C$ and $C^{\dagger}=C$, together with the corresponding Schr\"odinger-Robertson inequality
\begin{equation}
(\Delta A)^{2}(\Delta B)^{2}\geq \vert \frac{1}{4}\vert\langle C \rangle\vert^{2}+\sigma(A,B) \, , \quad \sigma(A,B):=\frac{\langle AB+BA\rangle}{2}-\langle A\rangle\langle B \rangle \, ,
\label{var1}
\end{equation}
with mean $\langle F \rangle:=\langle\Psi\vert F\vert\Psi\rangle$, variance $(\Delta F)^{2}=\langle F^{2}\rangle-\langle F \rangle^{2}$, and $\sigma(A,B)$, the \textit{correlation function}~\cite{Pur94}. Then, we say that $\vert\Psi\rangle$, with $\Vert \vert\Psi\rangle \Vert<\infty$, is a squeezed state if one of the variances associated to the observables $A$ and $B$ takes values below the uncertainty minimum $\sqrt{\frac{1}{4}\vert\langle C\rangle\vert^{2}+\sigma(A,B)}$ while the second variance compensates by increasing such that the inequality~\eqref{var1} is always saturated. Note that the definition of squeezing is in reference to the observable whose variance is being `squeezed'. 

Interestingly, if an underlying algebra can be identified with the observables $A$ and $B$,the squeezed states can be constructed by the sequential action of unitary operators on a fiducial state. Such unitary operators are usually constructed as the exponential representation of the algebra elements. To this end, there exist a great deal of examples of squeezed states in the literature such as the coherent squeezed states, coherent and squeezed number states~\cite{Nie97}, second-order squeezed states~\cite{Mar97}, and Susskind-Glogower coherent states~\cite{Leo11a,Gaz21b} (also know as London coherent states~\cite{Moy21}) to mention some. On the other hand, if a closed algebra is not available, one may proceed by solving an eigenvalue equation of the form $(A+\mathrm{i}\lambda B)\vert\psi\rangle=z\vert\psi\rangle$, which minimises~\eqref{var1}. See for instance~\cite{Alv02,Zel18,Zel21}. The latter constructions have been extended to quantised electromagnetic fields composed of several modes. Some examples include two-mode~\cite{Man97,Ger95,Thi15,Mor21} and higher-mode constructions \cite{glasserzhang}.

In this work, we exploit the overcompleteness of certain families squeezed states as a means of quantisation in two dimensions. The paper is structured as follows. In Sec.~\ref{sec:conv-ss} we review the basics of quantisation with one-mode squeezed states and define their semiclassical portraits. Following this, in Sec.~\ref{sec:SS2D} we generalise the notions of the preceding section to the two-mode case. We first define the most natural extension, the separable squeezed states, as the tensor product of two one-mode squeezed states acting on each mode independently, and then we define the non-separable squeezed states which cannot be factorised by a tensor product. We compare the quantisation of some classical functions in both cases and find that non-separability leads to mixing between quadrature operators between both modes. In Sec.~\ref{sec:PDM} we study the semiclassical portraits of a position dependent mass system in constrained geometry as an application, before concluding in Sec.~\ref{sec.con} with some remarks about extensions of the ideas presented in this paper to different problems.

\section{One-mode squeezed states quantisation}
\label{sec:conv-ss}
Before proceeding to the two-mode quantisation, let us recapitulate some results in one-dimensional squeezed state quantisation. Firstly, the unitary displacement and squeezing operators are defined as follows
\begin{equation}
D(\alpha):=e^{\alpha a^{\dagger}-\alpha^{*}a} \, , \quad S(\xi):=e^{-\frac{1}{2}\xi a^{\dagger 2}+\frac{1}{2}\xi^{*}a^{2}} \, , \quad \alpha,\xi\in\mathbb{C} \, ,
\label{DS1}
\end{equation}
in terms of the boson operators $a$ and $a^{\dagger}$, whose action on the elements of the elements of the Fock basis $\{\vert n\rangle\}_{n=0}^{\infty}$ is given by 
\begin{equation}
a\vert n+1\rangle = \sqrt{n+1}\vert n\rangle \, , \quad a^{\dagger}\vert n\rangle=\sqrt{n+1}\vert n+1\rangle \, \quad n=0,1\ldots \, , 
\end{equation}
along with the annihilation of the vacuum state, $a\vert0\rangle=0$. The squeezed coherent states, $\vert\alpha;\xi\rangle$, are then constructed through the action of the unitary operators~\eqref{DS1} on the corresponding fiducial state $\vert 0\rangle$,
\begin{equation}\label{ssdef}
 \vert\alpha;\xi\rangle=S(\xi)D(\alpha)\vert 0\rangle.
\end{equation}
Note that the alternative definition of squeezed states, $ \vert\xi;\beta\rangle=D(\beta)S(\xi)\vert 0\rangle$, is equivalent to \eqref{ssdef} through a braiding relation and this amounts to a relabelling of the parameters. Following the customary procedure, one can disentangle the unitary operators in the product of exponential functions in terms of $a$ and $a^{\dagger}$ separately. Alternatively, we can determine the eigenvalue equation related to $\vert\alpha,\xi\rangle$. This is achieved by computing the unitary transformations on the boson ladder operators
\begin{align}
S^{\dagger}(\xi)aS(\xi)=a\cosh\vert\xi\vert-a^{\dagger}\frac{\xi}{\vert\xi\vert}\sinh\vert\xi\vert \, , \quad D^{\dagger}(\alpha)aD(\alpha)=a+\alpha \, ,
\label{SaS}
\end{align}
where a Baker–Campbell–Hausdorff identity~\cite{Gri18} has been used. Thus, from the unitary transformation $D^{\dagger}S^{\dagger}aSD$, and after several calculations, we get the eigenvalue equation
\begin{equation}
(a+\tau a^{\dagger})\vert\alpha,\xi\rangle=\alpha\sqrt{1-\vert\tau\vert^{2}}\vert\alpha,\xi\rangle \, , \quad \tau=\frac{\xi}{\vert\xi\vert}\tanh\vert\xi\vert \, .
\end{equation}
The latter leads to a second-order finite-difference equation~\cite{Zel21d} in the Fock basis which yields the following normalisable states \cite{doi:10.1119/1.16337}
\begin{equation}
\vert\alpha;\xi\rangle=(1-\vert\tau\vert^{2})^{1/4}e^{-\frac{\vert\alpha\vert^{2}}{2}+\frac{\alpha^{2}\tau^{*}+\alpha^{*2}\tau}{4}}\sum_{n=0}^{\infty}\frac{\tau^{n/2}}{(2^{n}n!)^{1/2}}H_{n}\left(\alpha\sqrt{\frac{1-\vert\tau\vert^{2}}{2\tau}}\right)\, \vert n\rangle \, ,
\label{conv-ss-expansion}
\end{equation}
for $\alpha\in\mathbb{C}$ and $\vert\tau\vert<1$ ($\xi\in\mathbb{C}$). 

In general, the squeezed states do not form an orthogonal set of states as they have a non-zero overlap, $\langle\alpha',\xi\vert\alpha,\xi\rangle\neq 0$. Nevertheless they form an overcomplete set of states on the Hilbert space as they fulfil the resolution of the identity 
\begin{equation}
\int_{\alpha\in\mathbb{C}}\frac{\mathrm{d}^{2}\alpha}{\pi} \, \vert\alpha,\xi\rangle\langle\alpha,\xi\vert=\mathbb{I} \, , 
\label{conv-ss-measure}
\end{equation}
with $\mathbb{I}$ the identity operator in the Fock space $\mathcal{H}=\operatorname{span}\{\vert n\rangle\}_{n=0}^{\infty}$, and the measure function is uniform $\pi^{-1}$ as it is for the canonical coherent states.

From the very definition of the squeezed states, $\vert\alpha,\xi\rangle=S(\xi)D(\alpha)\vert 0\rangle$, we obtain a more simple form for the resolution of the identity, which reads
\begin{equation}
S(\xi)\left(\int_{\alpha\in\mathbb{C}}\frac{\mathrm{d}^{2}\alpha}{\pi} \,\vert\alpha\rangle\langle\alpha\vert\right)S^{\dagger}(\xi)=\mathbb{I} \, , \quad \vert\alpha\rangle=D(\alpha)\vert 0 \rangle \, , 
\label{conv-ss-ident1}
\end{equation}
with $\vert\alpha\rangle$ the conventional Glauber-Sudarshan coherent states. By defining the squeezed coherent states using the convention in \eqref{ssdef}, the measure function is constant. In this form, we have shown that squeezed states form an overcomplete set $\{\vert\alpha,\xi\rangle\}_{\alpha\in\mathbb{C}}$ with a uniform measure. The identity operator can be alternatively achieved through the orthogonality property related to the holomorphic Hermite polynomial~\cite{Eij90,Mou14}. See App.~\ref{sec:conv-ss-holH} for a detailed proof.

The resolution of the identity ensures that every element $\vert\phi\rangle\in\mathcal{H}$ can be expanded in the non-orthogonal basis $\{\vert\psi(\alpha)\rangle\}_{\alpha\in\mathbb{C}}$ through 
\begin{equation}
\vert \phi\rangle=\int_{\alpha\in\mathbb{C}}\frac{\mathrm{d}^{2}\alpha}{\pi}\, \mathcal{F}_{\phi}(\alpha)\vert\alpha,\xi\rangle \, , \quad \mathcal{F}_{\phi}(\alpha,\xi):=\langle\alpha,\xi\vert\phi\rangle \, ,
\end{equation}
where $\mathcal{F}_{\phi}(\alpha)$ is uniquely defined for each vector $\vert\phi\rangle$. 

Throughout this manuscript, we will use an alternative representation for the resolution of the identity~\eqref{conv-ss-measure} that encodes information about the position and momentum observables. To this end, let us recall the following relationships:
\begin{equation}
\hat{x}:=\lambda\frac{\hat{a}+\hat{a}^{\dagger}}{\sqrt{2}} \, , \quad \hat{p}:=\frac{\hbar}{\lambda}\frac{\hat{a}-\hat{a}^{\dagger}}{\mathrm{i}\sqrt{2}} \, ,
\label{quad-1d}
\end{equation}
where $\lambda>0$ is a free parameter with units of length. The latter can be alternatively defined through $\lambda=\hbar/\wp$, where $\wp$ is a free parameter with units of momentum. Such a definition is equivalent and can be used interchangeably. See~\cite{Gaz19,Gaz20a} for more details.

From~\eqref{quad-1d}, a relationship between the coherence parameter $\alpha=\operatorname{Re}\alpha+\mathrm{i}\operatorname{Im}\alpha$ and the expectation values $q\equiv\langle \hat{x}\rangle$ and $p\equiv\langle\hat{p}\rangle$ associated to the canonical position and momentum operators, respectively, with $\langle \cdot\rangle\equiv\langle\alpha,\xi\vert\cdot\vert\alpha,\xi\rangle$. By combining~\eqref{SaS} with~\eqref{quad-1d}, and averaging in the squeezed state basis we obtain the symplectic transform
\begin{equation}
\left(\begin{aligned} \operatorname{Re}[\alpha] \\ \operatorname{Im}[\alpha] \end{aligned}\right)=\left(\begin{alignedat}{3} &\frac{1+\operatorname{Re}[\tau]}{\sqrt{1-\vert\tau\vert^{2}}} \quad && \frac{\operatorname{Im}[\tau]}{\sqrt{1-\vert\tau\vert^{2}}} \\ & \frac{\operatorname{Im}[\tau]}{\sqrt{1-\vert\tau\vert^{2}}} \quad && \frac{1-\operatorname{Re}[\tau]}{\sqrt{1-\vert\tau\vert^{2}}} \end{alignedat} \right)\left(\begin{aligned} \frac{q}{\lambda\sqrt{2}} \\ \frac{\lambda p}{\hbar\sqrt{2}} \end{aligned}\right) \, .
\label{conv-ss-qp1}
\end{equation}
Note that, for $\tau=0$, we recover the well-known relationships $\operatorname{Re}[\alpha]=\frac{q}{\lambda\sqrt{2}}$ and $\operatorname{Im}[\alpha]=\frac{\lambda p}{\hbar\sqrt{2}}$ for coherent states. 

From~\eqref{conv-ss-qp1}, one may notice that $\alpha$ is linear in the expectation values $q=\langle \hat{x}\rangle$ and $p=\langle\hat{p}\rangle$. Thus, the complex-plane $\alpha$ can be understood as an analogue of the classical phase space manifold as every point $(q,p)\in\mathbb{R}^{2}$ is in unique correspondence with $\alpha\in\mathbb{C}$. Moreover, the transformation from the point $(q,p)$ to $\alpha$ given in~\eqref{conv-ss-qp1} is determined by a unimodular matrix, and thus the existence of the respective inverse transformation is guaranteed. The differential element in both frames is preserved, that is, $d^{2}\alpha\rightarrow (2\hbar)^{-1}\mathrm{d}q\mathrm{d}p$. With this identification, we can alternatively rewrite the resolution of the identity in terms of $q$ and $p$ as
\begin{equation}
\mathbb{I}=\int_{\mathbb{R}^{2}}\frac{\mathrm{d}q\mathrm{d}p}{2\pi\hbar}\vert q,p;\xi\rangle\langle q,p;\xi\vert \, , \quad \vert q,p;\xi\rangle\equiv\vert\alpha(q,p);\xi\rangle \,  ,
\label{conv-ss-ident-2}
\end{equation}
with $\alpha(q,p)$ given in~\eqref{conv-ss-qp1}. 

\begin{figure}
\centering
\subfloat[][$\operatorname{Re}\left(\sigma_{q}^{2}\right)$]{\includegraphics[width=0.3\textwidth]{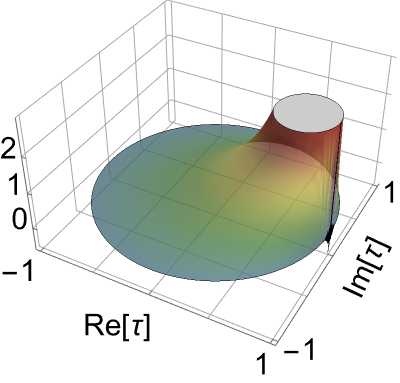}
\label{fig:varmass1-1aaaa}}
\hspace{20mm}
\subfloat[][$\Delta_p^2$]{\includegraphics[width=0.3\textwidth]{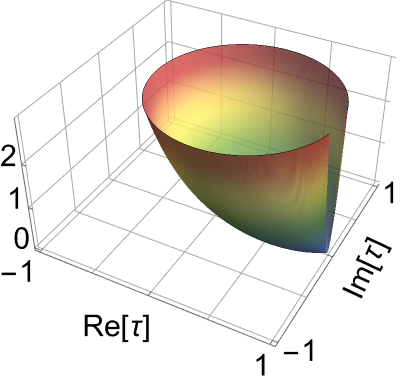}
\label{fig:varmass1-2aaaa}}
\caption{Real part of $\sigma_{q}^{2}$ given in~\eqref{sigma} and $\Delta_p^2$ given in \eqref{conv-ss-semic2} in terms of the real and imaginary parts of $\tau$, for $\vert\tau\vert<1$.}
\label{fig:sigma}
\end{figure}

It is useful to determine the position representation for the squeezed states, $\psi_{\{q,p;\xi\}}(x)=\langle x\vert q,p;\xi\rangle$, as it facilitates the determination of some observables. From~\eqref{conv-ss-expansion}, together with $\langle x\vert n\rangle=(2^{n}n!\sqrt{\pi})^{-\frac{1}{2}}e^{-\frac{x^{2}}{2\lambda^{2}}}H_{n}(\frac{x}{\lambda})$, and using the summation identities for Hermite polynomials~\cite{Pru86} we obtain the normalised wavefunction
\begin{equation}
\psi(\alpha;\xi;x)(x):=\langle x\vert\alpha;\xi\rangle=\frac{(1-\vert\tau\vert^{2})^{1/4}}{\pi^{1/4}\sqrt{1-\tau}}e^{-\frac{\vert\alpha\vert^{2}}{2}}e^{\frac{\alpha^{2}\tau^{*}+\alpha^{*2}\tau}{4}}e^{-\frac{\alpha^{2}(1-\vert\tau\vert^{2})}{2(1-\tau)}} \, e^{-\frac{1}{2}\left(\frac{1+\tau}{1-\tau}\right)\frac{x^{2}}{\lambda^{2}}}e^{\frac{\sqrt{2(1-\vert\tau\vert^{2})}\alpha}{1-\tau}\frac{x}{\lambda}} \, .
\label{ss-wf}
\end{equation}
Alternatively, we can rewrite \eqref{ss-wf} in terms of the expectation values $q$ and $p$ by using the relationships~\eqref{conv-ss-qp1} to get, up to a complex-phase,
\begin{equation}
\psi(q,p;\xi,x)=\frac{1}{\pi^{1/4}\sqrt{\lambda}}\frac{(1-\vert\tau\vert^{2})^{1/4}}{\vert 1-\tau\vert^{1/2}}\exp\left(-\frac{\sigma^{2}_{q}}{2\lambda^{2}}(q-x)^{2}+\mathrm{i}\frac{\operatorname{Im}[\tau]}{4\lambda^{2}}\left(q-\frac{\lambda^{2}}{\hbar}p\right)^{2}+\mathrm{i}\frac{p}{\hbar}\left(x-\frac{q}{2}\right)\right) \, ,
\label{conv-ss-wf}
\end{equation}
where $\sigma_{q}^{2}$ is a complex parameter given by
\begin{equation}
\sigma_{q}^{2}:=\frac{(1-\vert\tau\vert^{2})+2\mathrm{i}\operatorname{Im}[\tau]}{\vert 1-\tau\vert^{2}} \, .
\label{sigma}
\end{equation}
This parameter diverges for $\tau\rightarrow 1$, which is excluded from the domain $\tau\in\vert\tau\vert<1$. The real part of $\sigma_{q}^{2}$ is a positive definite function in such a domain, so that~\eqref{conv-ss-wf} describes a well-defined Gaussian function, where $2\lambda^{2}\operatorname{Re}[\sigma_{q}^{-2}]$ plays the role of the Gaussian width. The behaviour of Re$[\sigma_{q}^{2}]$ is depicted in Fig.~\ref{fig:sigma} as a function of the real and imaginary parts of $\tau$ inside the complex unit-disk. It is clear that the Gaussian wavepacket squeezes in the vicinity of $\tau=1$.

\subsubsection*{Quantisation}
We now proceed to discuss one of the main results of this manuscript, the quantisation map using squeezed states. Although we summarise the results from the one-dimensional case, the results developed here extend to higher dimensions with relative ease. To this end, let us introduce an operation that maps a classical observable $f(\alpha) \equiv f(q,p)$, defined in the classical phase space manifold, into a linear operator $\hat{A}_{f}$, defined to act on elements of the vector space $\mathcal{H}$. This a procedure is known as a \textit{quantisation map}, which requires a complete family of states, like the squeezed states, such that to every classical observable we can associate a unique quantum observable. The map is defined by
\begin{equation}
f(q,p)\mapsto \hat{A}_{f}:=\int_{\mathbb{R}^{2}}\frac{\mathrm{d}q\mathrm{d}p}{2\pi \hbar}f(q,p)\vert q,p;\xi\rangle\langle q,p;\xi\vert \, .
\label{conv-ss-qmap}
\end{equation}
Although this definition is quite general, in some cases it can be computationally infeasible. To overcome this issue, we take advantage of the coordinate representation in order to compute the action of the observable $A_{f}$ on a ``test function,'' which is an arbitrary element $\vert\Psi\rangle\in\mathcal{H}$. This corresponds to the operation
\begin{equation}
\left(A_{f}^{(op)}\Psi\right)(x) \equiv \langle x \vert \hat{A}_{f} \vert \Psi\rangle=\int_{\mathbb{R}} \mathrm{d}x'\, \mathcal{K}_{f}(\xi;x,x') \Psi(x') \, , 
\label{conv-ss-kernel1}
\end{equation}
where $A_{f}^{(op)}$ is the coordinate representation of $\hat{A}_{f}$, together with $\mathcal{K}_{f}(\xi;x,x')$, a kernel operator containing information about the action of $\hat{A}_{f}$ on the test function $\Psi(x)$, determined through
\begin{equation}
\mathcal{K}_{f}(\xi;x,x'):=\int_{\mathbb{R}^{2}}\frac{\mathrm{d}q\mathrm{d}p}{2\pi\hbar}f(q,p)\psi^{*}(q,p;\xi;x') \psi(q,p;\xi;x) \, ,
\label{conv-ss-kernel2}
\end{equation}
with $\psi(q,p;\xi;x)$ the wavefunction given in~\eqref{conv-ss-wf}.

To illustrate the use of the kernel representation~\eqref{conv-ss-kernel1}, we consider two examples.

$\bullet$ First, let $f(q,p)=q$ such that the kernel becomes $\mathcal{K}_{q}(\xi;x,x')=\delta(x'-x)$, where we have used some elementary properties of the Fourier transform while integrating with respect to $p$. In this form, we get $\left(A_{q}^{(op)}\Psi\right)(x)\equiv x\Psi(x)$, which means that $q\mapsto\hat{A}_{q}=\hat{x}\equiv x$, as expected. 

$\bullet$ Similarly, for $f(q,p)=p$, and using some properties involving derivatives of the Fourier transform, we obtain the kernel $\mathcal{K}_{q}(\xi;x,x')=-\mathrm{i}\hbar\delta_{x'}(x'-x)$, with the subscript index denoting the partial derivative with respect to $x'$. Such a kernel leads to $\left(A_{p}^{(op)}\Psi\right)(x)\equiv \frac{\hbar}{\mathrm{i}}\frac{\partial}{\partial x}\Psi(x)$. That is, the quantisation of $p$ becomes, in the $x$-representation, proportional to the derivative with respect to $x$, $p\mapsto\hat{A}_{p}=\hat{p}\equiv\frac{h}{\mathrm{i}}\frac{\partial}{\partial x}$.

$\bullet$ The quantisation of $f(q,p)=qp$ follows from the previous two cases, leading to $\hat{A}_{qp}=\frac{\hat{x}\hat{p}+\hat{p}\hat{q}}{2}-\frac{\operatorname{Im}[\sigma_{q}^{2}]}{2\operatorname{Re}[\sigma_{q}^{2}]}$. This corresponds to the symmetrisation rule of the operator product of $\hat{x}$ and $\hat{p}$ plus a constant term that depends explicitly on the squeezing parameter. For $\tau\in\mathbb{R}$, we recover the conventional symmetrisation rule. Moreover, the operator $\hat{A}_{qp}$ is the generator of dilations for $\hat{x}$ and $\hat{p}$. Explicitly, the unitary operator $U_{d}(\ell):=e^{\mathrm{i}\ell\hat{A}_{qp}}$ induces the unitary transformations $U_{d}(\ell)\hat{x}U_{d}^{\dagger}(\ell)=e^{\ell}\hat{x}$ and $U_{d}(\ell)\hat{p}U_{d}^{\dagger}(\ell)=e^{-\ell}\hat{p}$.

Before proceeding, it is worth mentioning that the definition~\eqref{conv-ss-qmap} fulfills two fundamental properties required by any quantisation mechanism~\cite{Cre89,Gos16}. Firstly, the quantisation map~\eqref{conv-ss-qmap} must promote the classical function $f(q,p)=1$ into the identity operator $\mathbb{I}$. This is already guaranteed from the completeness relationship~\eqref{conv-ss-ident-2}. Secondly, Dirac's correspondence rule should be recovered, $\{q,p\}_{PB}=1\rightarrow [\hat{x},\hat{p}]=i\hbar$, with $\{f(q,p),h(q,p)\}_{PB}$ the Poisson brackets~\cite{Gol11}. From the previous two examples, it follows directly that $[A_{q}^{(op)},A_{p}^{(op)}]\Psi(x)=\mathrm{i}\hbar\Psi(x)$, which fulfils the correspondence rule.

\subsubsection*{Semiclassical portraits}
Interestingly, as with the conventional coherent states, we can define a set of quantities that behave analogously to their classical counterparts. These quantities are known as \textit{semiclassical portraits}~\cite{Gaz19,Gaz20a}, which are defined as the expectation values of the corresponding quantum observables $\hat{A}_{f}$ in the squeezed states basis. We thus introduce the \textit{lower symbol}, or semiclassical portrait, as
\begin{equation}
\begin{aligned}
f(q,p)\mapsto \widecheck{A}_{f}:=&\langle q,p;\xi\vert \hat{A}_{f} \vert q,p;\xi\rangle=\int_{\mathbb{R}^{2}}\frac{\mathrm{d}q\mathrm{d}p}{2\pi \hbar}f(q',p')\vert\langle q',p';\xi\vert q,p;\xi\rangle\vert^{2}\, .
\end{aligned}
\label{conv-ss-semic1}
\end{equation}
where the absolute value square overlap between squeezed states is given by
\begin{equation}
\vert\langle q',p';\xi\vert q,p;\xi\rangle\vert^{2}=e^{-\frac{\Delta_{q}^{2}}{2\lambda^{2}}(q-q')^{2}-\frac{\lambda^{2}}{2\hbar^{2}}\Delta_{p}^{2}(p-p')^{2}-2\frac{\gamma}{\hbar}(q-q')(p-p')} \, ,
\label{overlap-SS}
\end{equation}
with the Gaussian widths $\Delta_{q}$ and $\Delta_{p}$, together with the coupling parameter $\gamma$, given in terms of the original parameters by
\begin{equation}
\begin{aligned}
&\Delta_{q}^{2}:=\vert\sigma_{q}\vert^{2}=\frac{1-\vert\tau\vert^{2}}{\vert 1-\tau\vert^{2}}+\frac{4\operatorname{Im}[\tau]^{2}}{\vert 1-\tau\vert^{2}(1-\vert\tau\vert^{2})} \, , \quad \Delta_{p}^{2}=\frac{\vert 1-\tau\vert^{2}}{1-\vert\tau\vert^{2}} \, , \quad \gamma=\frac{\operatorname{Im}[\tau]}{1-\vert\tau\vert^{2}} \, .
\end{aligned}
\label{conv-ss-semic2}
\end{equation}

A handy formula can be derived for classical functions that depend only on position, $f(q,p)=h(q)$, in which case the integral~\eqref{conv-ss-semic1} becomes
\begin{equation}
\widecheck{A}_{h(q)}=\frac{1}{\sqrt{2\pi}\lambda \Delta_{p}}\int_{\mathbb{R}}\mathrm{d}q' h(q') e^{-\frac{(q-q')^{2}}{2\lambda^{2}\Delta_{p}^{2}}} \, .
\label{1D-hq}
\end{equation}
This may be thought of as a Gaussian regularisation of the classical function $h(q)$. This is particularly useful when dealing with discontinuous functions $h(q)$. Further examples will be discussed once we introduce the two-mode extension in the upcoming sections.

\section{Families of two-mode squeezed states}
\label{sec:SS2D}
We now turn our attention to the main purpose of the paper: quantisation for two-dimensional systems. As we discussed in Sec.~\ref{sec:conv-ss}, the quantisation map depends on the choice of the family of overcomplete states used, and thus the quantisation for two-dimensional systems can be constructed in a similar vein to the one-dimensional case by implementing families of multimode states, like multimode coherent states, such that they fulfil the resolution of the identity. Here, we define the larger Hilbert space $\mathcal{H}=\operatorname{span}\{\vert n_{1}\rangle\otimes\vert n_{2}\rangle \}_{n_{1},n_{2}=0}^{\infty}$, with $\vert n_{j}\rangle$ elements of the Fock basis on the $j$ mode, and the corresponding identity operator in such a vector space reads as $\mathbb{I}=\sum_{n_{1},n_{2}=0}^{\infty}\vert n_{1}, n_{2}\rangle\langle n_{1},n_{2}\vert$, with $\vert n_{1},n_{2}\rangle\equiv \vert n_{1}\rangle\otimes\vert n_{2}\rangle$.
In this form, we introduce the set of multimode bosonic operators $\{ a_{1}, a_{1}^{\dagger},a_{2},a_{2}^{\dagger}\}$, which fulfil the commutation relationships $[a_{j},a_{k}^{\dagger}]=\delta_{j,k}$, for $j,k=1,2$. Moreover, the action of such multimode operators on the extended vector space $\mathcal{H}$ is defined as
\begin{equation}
\begin{aligned}
& a_{1}\vert n_{1},n_{2}\rangle=\sqrt{n_{1}}\vert n_{1}-1,n_{2}\rangle \, , \quad a_{2}\vert n_{1},n_{2}\rangle=\sqrt{n_{2}}\vert n_{1},n_{2}-1\rangle \, , \\
& a_{1}^{\dagger}\vert n_{1},n_{2}\rangle=\sqrt{n_{1}+1}\vert n_{1}+1,n_{2}\rangle \, , \quad a_{2}^{\dagger}\vert n_{1},n_{2}\rangle=\sqrt{n_{2}+1}\vert n_{1},n_{2}+1\rangle \, .
\label{multi-boson}
\end{aligned}
\end{equation}
The canonical canonical position and momentum quadratures $\hat{x}_{j}$ and $\hat{p}_{j}$, respectively, for the $j$ mode are related to the multimode boson operators as
\begin{equation}
\hat{x}_{j}:=\lambda_{j}\frac{a_{j}+a_{j}^{\dagger}}{\sqrt{2}} \, , \quad \hat{p}_{j}:=\frac{\hbar}{\lambda_{j}}\frac{a_{j}-a_{j}^{\dagger}}{\mathrm{i}\sqrt{2}} \, , \quad j=1,2 \, ,
\label{2d-quad}
\end{equation}
where $[\hat{x}_{j},\hat{p}_{k}]=\mathrm{i}\hbar\delta_{j,k}$ and $[\hat{x}_{j},\hat{x}_{k}]=[\hat{p}_{j},\hat{p}_{k}]=0$, for $j,k=1,2$. 

Throughout this section we focus on two particularly interesting cases. The first one being the most immediate extension by taking the tensor product of two independent one-mode squeezed states. In the second case, we consider a family of two-mode states that do not factorize as the tensor product of two one-mode squeezed states.

\subsection{Separable two-dimensional squeezed states}
\label{sec:sep-squeezed-quantization}
Let us consider the conventional one-mode squeezed states introduced in Sec.\ref{sec:conv-ss}, and extend them into the extended vector space $\mathcal{H}$ through the direct product of two squeezed, one in each mode, with the coherence and squeezing parameters in general being different in each mode. Henceforth, we refer to this specific construction as \textit{separable squeezed states}, which are explicitly defined as
\begin{equation}
\vert \vec{\alpha};\vec{\xi}\rangle = \vert\alpha_{1},\xi_{1}\rangle\otimes\vert\alpha_{2};\xi_{2}\rangle=S(\xi_{2})S(\xi_{1})D(\alpha_{2})D(\alpha_{1})\vert 0,0 \rangle \, , \quad \alpha_{j},\xi_{j}\in\mathbb{C} \, , \quad j=1,2 \, ,
\label{2d-squeezed}
\end{equation}
with $D(\alpha_{j})$ and $S(\xi_{j})$ denoting the displacement and squeezing operators, respectively, defined on the $j$ mode, and $\vert 0,0\rangle$ the two-mode vacuum. 

From the separable squeezed states~\eqref{2d-squeezed}, we can find a relationship between the coherent parameter and the expectation value of the canonical coordinates. This is done analgously to the one-mode case, and we find
\begin{equation}
\begin{pmatrix}
\vec{r}_{{\alpha}_{1}} \\
\vec{r}_{{\alpha}_{2}}
\end{pmatrix}
=
\begin{pmatrix}
\mathbb{M}_{1} & \mathbb{O} \\
\mathbb{O} &  \mathbb{M}_{2}
\end{pmatrix} 
\begin{pmatrix}
\vec{r}_{1} \\
\vec{r}_{2}
\end{pmatrix}
\, , \quad 
\mathbb{M}_{j}=
\begin{pmatrix}
\frac{1+\operatorname{Re}[\tau_{j}]}{\sqrt{1-\vert\tau_{j}\vert^{2}}} \quad & \frac{\operatorname{Im}[\tau_{j}]}{\sqrt{1-\vert\tau_{j}\vert^{2}}} \\
\frac{\operatorname{Im}[\tau_{j}]}{\sqrt{1-\vert\tau_{j}\vert^{2}}} \quad & \frac{1-\operatorname{Re}[\tau_{j}]}{\sqrt{1-\vert\tau_{j}\vert^{2}}}
\end{pmatrix} 
\, , \quad \tau_{j}=\frac{\xi_{j}}{\vert\xi_{j}\vert}\tanh\vert\xi_{j}\vert \, , 
\label{sep-alpha-qp}
\end{equation}
with $\vert\xi_{j}\vert<1$ and $j=1,2$. $\mathbb{O}$ stands for the null $2\times 2$ matrix and
\begin{equation}\label{parametertransf}
\vec{r}_{{\alpha}_{j}}=
\begin{pmatrix}
\operatorname{Re}[\alpha_{j}] \\
\operatorname{Im}[\alpha_{j}]
\end{pmatrix} 
\, , \quad  
\vec{r}_{j}=
\begin{pmatrix}
\frac{q_{j}}{\lambda_{j}\sqrt{2}} \\ 
\frac{\lambda_{j}p_{j}}{\hbar \sqrt{2}}
\end{pmatrix}
\, , \quad
q_{j}=\langle \hat{x}_{j}\rangle \, , \quad p_{j}=\langle \hat{p}_{j}\rangle \, .
\end{equation}

We remark that the limit $\tau_j \rightarrow \infty$ refers to infinite squeezing in the $j$-th mode. The separable squeezed states minimise the Schr\"odinger-Robertson uncertainty relation for the physical position and momentum quadratures in each mode independently. That is, 
\begin{equation}
(\Delta \hat{x}_{j})^{2}(\Delta \hat{p}_{j})^{2}=\frac{\hbar^{2}}{4}+\tilde{\sigma}(\hat{x}_{j},\hat{p}_{j}) \, , \quad j=1,2 \, .
\end{equation}

Additionally, the separable squeezed states admit a coordinate representation defined in terms of the eigenstates of the quadratures $\hat{x}_{1}$ and $\hat{x}_{2}$ in a similar manner to their one-dimensional counterparts. By considering the linear transformation~\eqref{sep-alpha-qp}, we rewrite the squeezed states in terms of $q_{j}$ and $p_{j}$ so that the normalised wavefunction
$\psi(\vec{q},\vec{p};\vec{\xi},\vec{x}):=\langle \vec{x} \vert \vec{q},\vec{p};\vec{\xi} \rangle$, with $\vert\vec{x}\rangle=\vert x_{1}\rangle\otimes\vert x_{2}\rangle$, takes the form
\begin{equation}
\psi(\vec{q},\vec{p};\vec{\xi};\vec{x}):=\frac{\exp\left(-\frac{1}{2}\sum_{j=1}^{2}\left[\frac{\sigma_{q_{j}}^{2}}{\lambda_{j}^{2}}(q_{j}-x_{j})^{2}-\mathrm{i}\frac{\operatorname{Im}[\tau_{j}]}{2\lambda_{j}^{2}}\left(q_{j}-\frac{\lambda_{j}^{2}}{\hbar}p_{j}\right)^{2}\right]+\mathrm{i}\frac{\vec{p}}{\hbar}\cdot\left(\vec{x}-\frac{\vec{q}}{2}\right)\right)}{\sqrt{\lambda_{1}\lambda_{2}\Delta_{p_{1}}\Delta_{p_{2}}}\pi^{1/4}} \, , 
\end{equation}
where
\begin{equation}
\sigma_{q_{j}}^{2}:=\frac{(1-\vert\tau_{j}\vert^{2})+2\mathrm{i}\operatorname{Im}[\tau_{j}]}{\vert 1-\tau_{j}\vert^{2}} \, , \quad \Delta_{p_{j}}^{2}=\frac{\vert 1-\tau_{j}\vert^{2}}{1-\vert\tau_{j}\vert^{2}} \, , \quad j=1,2 \, .
\end{equation}

\subsubsection*{Quantisation map and semiclassical portraits}
In Sec.~\ref{sec:conv-ss}, we showed that the one-mode squeezed states form an overcomplete family of states. This property is inherited by the two-dimensional case in the extended vector space $\mathcal{H}$ through
\begin{equation}
\mathbb{I}=\int_{\mathbb{R}^{4}}\frac{\mathrm{d}^{2}\vec{q}\mathrm{d}^{2}\vec{p}}{(2\pi\hbar)^{2}}\vert\vec{q},\vec{p};\vec{\xi}\rangle\langle\vec{q},\vec{p};\vec{\xi}\vert \, , \quad \mathrm{d}^{2}\vec{q}=\mathrm{d}q_{1}\mathrm{d}q_{2} \, , \quad \mathrm{d}^{2}\vec{p}=\mathrm{d}p_{1}\mathrm{d}p_{2} \, .
\end{equation}
The latter can be easily shown by factorising the separable squeezed states into its independent modes and then using the corresponding one-mode results.

In this form, the quantisation map is implemented straightforwardly through the integral transform
\begin{equation}
f(\vec{q},\vec{p})\mapsto \hat{A}_{f}:=\int_{\mathbb{R}^{4}}\frac{\mathrm{d}\vec{q}\mathrm{d}\vec{p}}{(2\pi\hbar)^{2}}f(\vec{q},\vec{p})\vert\vec{q},\vec{p};\vec{\xi}\rangle\langle\vec{q},\vec{p};\vec{\xi}\vert \, ,
\end{equation}
which can be conveniently rewritten in terms of the coordinate representation as
\begin{equation}
\left(A^{(op)}_{f}\Psi\right)(\vec{x})=\int_{\mathbb{R}^{2}}\mathrm{d}\vec{x}\,' \mathcal{K}_{f}(\vec{\xi};\vec{x},\vec{x}\,')\Psi(\vec{x}\,') \, ,
\end{equation}
where the integral kernel is given by
\begin{equation}
\mathcal{K}_{f}(\xi;\vec{x},\vec{x}\,'):=\int_{\mathbb{R}^{2}}\frac{\mathrm{d}\vec{q}\mathrm{d}\vec{p}}{2\pi\hbar}f(\vec{q},\vec{p})\psi^{*}(\vec{q},\vec{p};\vec{\xi};\vec{x}\,') \psi(\vec{q},\vec{p};\vec{\xi};\vec{x}) \, .
\label{sep-kernel2}
\end{equation}

Since the squeezed states are the tensor product of two one-dimensional states, the quantisation of a classical function of the form $f(\vec{q},\vec{p})=f_{1}(q_{1},p_{1})f_{2}(q_{2},p_{2})$ produces an operator factorisable as $f(q,p)\mapsto\hat{A}_{f_{1}}\otimes\hat{A}_{f_{2}}$. In particular, for a classical function $f(\vec{q},\vec{p})=h_{1}(q_1)h_{2}(q_2)$ we obtain a simplified kernel of the form
\begin{equation}
\mathcal{K}_{h_1 h_2}(\vec{\xi};\vec{x},\vec{x}\,')=\mathcal{K}_{h_{1}}(\xi_{1};x_{1},x_{1}')\mathcal{K}_{h_{2}}(\xi_{2};x_{2},x_{2}') \, ,
\end{equation}
with
\begin{equation}
\mathcal{K}_{h_{j}}(\xi_{j};x_{j},x_{j}'):=\delta(x_{j}-x_{j}')\frac{(1-\vert\tau_{j}\vert^{2})^{\frac{1}{2}}}{\pi^{\frac{1}{2}}\lambda_{j}\vert 1-\tau_{j}\vert} \int_{\mathbb{R}} \mathrm{d}q_{j}\, h_{j}(q_{j})e^{-\frac{\operatorname{Re}[\sigma_{q_{j}}^2]}{\lambda_{j}}(q_{j}-x_{j})^2} \, , \quad j=1,2 \, .
\label{sep-kernel4}
\end{equation}

In a similar vein, the construction of the corresponding semiclassical portraits follows straightforwardly from the one-dimensional case. That is, by averaging the quantised operators $\hat{A}_{f}$ over the two-mode separable squeezed state basis we get
\begin{equation}
\widecheck{A}_{f(\vec{q},\vec{p})}=\int_{\mathbb{R}^{4}}\frac{\mathrm{d}\vec{q}\,' \mathrm{d}\vec{p}\, '}{(2\pi\hbar)^{2}}f(\vec{q}\,',\vec{p}\,') \, \vert\langle \vec{q}\,',\vec{p}\,';\vec{\xi}\vert\vec{q},\vec{p};\vec{\xi}\rangle\vert^{2}
\end{equation}
where the squeezed state overlap is defined as the product of two one-dimensional squeezed states overlap given in~\eqref{overlap-SS}. In this form, we may distinguish the following cases:

$\bullet$ The semiclassical portrait of $f(\vec{q},\vec{p})=h(\vec{q})$ leads to

\begin{equation}
\widecheck{A}_{h(\vec{q})}=\frac{1}{2\pi\lambda_{1}\lambda_{2}\Delta_{p_{1}}\Delta_{p_{2}}}\int_{\mathbb{R}^{2}}\mathrm{d}q_{1}'\mathrm{d}q_{2}'h(\vec{q}\,')\exp\left( \vec{v'}\cdot \mathfrak{S} \vec{v'}\right) ,
\end{equation}
where the vector $\vec{v'}=\begin{pmatrix}
q_1-q_{1}' \\
q_2-q_{2}'
\end{pmatrix}$ and scaling matrix $\mathfrak{S}=\begin{pmatrix}
\frac{-1}{2\lambda_{1}^{2}[\Delta_{p_{1}}]^{2}} & 0 \\
0 & \frac{-1}{2\lambda_{2}^{2}[\Delta_{p_{2}}]^{2}}
\end{pmatrix}$. This is a Gaussian regularisation of the classical observables, analagously to the kernel regularisation obtained in~\eqref{sep-kernel4}.

$\bullet$ A general expression can be found if we consider a classical function that mixes one of the momenta with an arbitrary function of both positions, $f(\vec{q},\vec{p})=p_{j}h(\vec{q})$. In this setup we get
\begin{equation}
\widecheck{A}_{p_{j}h(\vec{q})}=p_{j}\widecheck{A}_{h(\vec{q})}+\frac{2\hbar\gamma_{j}}{\Delta_{p_{j}}^{2}\lambda_{j}^{2}} \left( q_{j}\widecheck{A}_{h(\vec{q})} - \widecheck{A}_{q_{j}h(\vec{q})} \right) \, , \quad j=1,2 \, .
\label{sep-semi-p1h}
\end{equation}
Clearly, for $h(\vec{q})=1$, we recover the expected result $\widecheck{A}_{p_{j}}=p_{j}$.

$\bullet$ From the previous two examples, we may compute the semiclassical portrait related to a kinetic energy of the form $f(\vec{q},\vec{p})=p_{j}^{2}h(\vec{q})$, with $h(\vec{q})=(m(\vec{q}))^{-1}$ playing the role of a position-dependent mass term. We obtain
\begin{multline}
\widecheck{A}_{p_{j}^{2}h(\vec{q})}=\left(p_{j}^{2}+\frac{\hbar^{2}}{\Delta_{p_{j}}^{2}\lambda_{j}^{2}}\right)\widecheck{A}_{h(\vec{q})}+\frac{4\hbar^{2}\gamma_{j}^{2}}{\Delta_{p_{j}}^{4}\lambda_{j}^{4}}\left(q_{j}^{2}\widecheck{A}_{h(\vec{q})}-2q_{j}\widecheck{A}_{q_{j}h(\vec{q})} +\widecheck{A}_{q_{j}^{2}h(\vec{q})}\right) + \\
\frac{4\hbar\gamma_{j}}{\Delta_{p_{j}}^{2}\lambda_{j}^{2}}p_{j}\left(q_{j}\widecheck{A}_{h(\vec{q})}-\widecheck{A}_{q_{j}h(\vec{q})}\right) \, ,
\label{sep-semi-p2h}
\end{multline}
for $j=1,2$. Note that the kinetic term is composed of $p_{j}^{2}\widecheck{A}_{h(\vec{q})}$, which includes the regularised semiclassical function associated with $h(\vec{q})$, plus terms proportional to $\hbar^{2}$ and $\hbar$. The latter induce the quantum effects resulting from the squeezed state quantisation and become relevant whenever $\lambda$ and $q_{j}$ are around the same order of magnitude as $\hbar$ (small-scale). That is, the semiclassical model still accounts for quantum effects in the small-scale, whereas quantum effects are negligible on the macroscopic scale ($\lambda,q_{j}>>\hbar$).

\subsection{Non-separable two-mode squeezed states}
\label{sec:non-sep-quantization}
The two-dimensional construction of squeezed states discussed in Sec.~\eqref{sec:sep-squeezed-quantization} is the most immediate generalisation of the one-dimensional squeezed states. However, those states are a particular extension, and in multidimensional systems more general states can be constructed which cannot be decomposed into the tensor product of one-dimensional states. Such classes of states have been discussed in the literature for the two-dimensional case by using two-mode ladder operators so that the information of both modes is mixed~\cite{Man97,Mor21}. 

In this section we follow the construction introduced in~\cite{Man97}, where a family of two-mode squeezed states are constructed with the aid of the mixing operator
\begin{equation}
U_{BS}(\phi):=e^{\phi(a^{\dagger}_1\otimes a_2-a_1\otimes a_2^{\dagger})} \, , \quad \phi\in[0,2\pi) \, ,
\end{equation}
which is equivalent to the quantum representation of the \textit{beam-splitter}. In what follows we will suppress the tensor product notation and it will be implicit that the operators labelled $a_1$ and $a_2$ act on the first and second modes, respectively. In this form, we may combine the beam-splitter with the one-mode  displacement and squeezing operators $D(\alpha_{j})$ and $S(\xi_{j})$, respectively, in order to construct the \textit{non-separable squeezed states}~\cite{Man97}
\begin{equation}
\vert\vec{\alpha};\vec{\xi}, \phi \rangle = G \vert 0 , 0 \rangle \, , \quad \vec{\alpha}=(\alpha_{1},\alpha_{2}) \, , \quad \vec{\xi}=(\xi_{1},\xi_{2}) \, .
\label{nonsep-SS}
\end{equation}
with $G$ the unitary operator
\begin{equation}
G = D(\alpha_{1})D(\alpha_{2})U_{BS}(\phi) S(\xi_{1})S(\xi_{2})  \, , \quad  \alpha_{j},\xi_{j}\in\mathbb{C} \, , \quad j=1,2 \, .
\label{G}
\end{equation}
The order of the displacement and squeezing operators has been deliberately chosen so that the squeezing operators act first on the two-mode vacuum state $\vert 0,0\rangle$. This is due the fact that the beam-splitter operator acting on a nonclassical state, such as the two-mode squeezed vacuum, produces a non-separable state at the output. Therefore, if we were to act with the displacement operator first, we would get a separable state at the output, as the coherent states are classical in this respect. See~\cite{Kim02} for details. In this form, the non-separability of the two-mode intertwined squeezed states is determined by the parameter $\phi$. For $\phi=0$, we recover the separable states of Sec.\ref{sec:conv-ss}.

Now, from~\eqref{G}, we can find the unitary transformation of the boson ladder operators for both the modes, $a_1$ and $a_2$, respectively.  We make use of the well-known Bogoliubov transformations~\cite{Nie97b} to obtain
\begin{equation}
\begin{aligned}
& G^{\dagger}a_{1}G=\alpha_{1}+\cos\phi\left(a_{1}\cosh\vert\xi_{1}\vert-a_{1}^{\dagger}\frac{\xi_{1}}{\vert\xi_{1}\vert}\sinh\vert\xi_{1}\vert\right)+\sin\phi\left(a_{2}\cosh\vert\xi_{2}\vert-a_{2}^{\dagger}\frac{\xi_{2}}{\vert\xi_{2}\vert}\sinh\vert\xi_{2}\vert\right) \, , \\
& G^{\dagger}a_{2}G=\alpha_{2}+\cos\phi\left(a_{2}\cosh\vert\xi_{2}\vert-a_{2}^{\dagger}\frac{\xi_{2}}{\vert\xi_{2}\vert}\sinh\vert\xi_{2}\vert\right)-\sin\phi\left(a_{1}\cosh\vert\xi_{1}\vert-a_{1}^{\dagger}\frac{\xi_{1}}{\vert\xi_{1}\vert}\sinh\vert\xi_{1}\vert\right) \, .
\end{aligned}
\label{Unit2D1}
\end{equation}
From the latter it is evident that $G$ indeed mixes the modes $a_{1}$ and $a_{2}$ where, for $\phi=0$, the transformation decouples $a_{1}$ from $a_{2}$. The unitary transformation~\eqref{Unit2D1}, combined with the definition of the physical canonical quadratures~\eqref{2d-quad}, allows us to recover the same relationships between the canonical coordinates and complex parameters $\alpha_i$ as in \eqref{parametertransf}.

The unitary transformations~\eqref{Unit2D1} lead to a set of two eigenvalue equations whose eigenfunctions are the two-dimensional squeezed states $\vert\vec{\alpha};\vec{\xi},\phi\rangle$, see App.~\ref{sec:nonsep-WF} for details. In this form, we obtain the corresponding wavefunction as
\begin{equation}
\psi(\vec{q},\vec{p};\vec{\xi},\phi;\vec{x}):= \mathcal{N}(\vec{q},\vec{p};\vec{\xi},\phi) \, e^{-\frac{\Delta_{1}}{\lambda_{1}^{2}}x_{1}^{2}-\frac{\Delta_{2}}{\lambda_{2}^{2}}x_{2}^{2}-\frac{\ell}{\lambda_{1}\lambda_{2}} x_{1}x_{2}+\frac{\ell_{1}}{\lambda_1}x_{1}+\frac{\ell_{2}}{\lambda_2}x_{2}} \, ,
\end{equation}
where $\mathcal{N}(\vec{q},\vec{p};\vec{\xi},\phi)$ is a normalisation factor, and the coefficients proportional to the bilinear terms in $x_{1}$ and $x_2$ are
\begin{equation}
\begin{aligned}
& \Delta_{1}:=\frac{1-\tau_{1}\tau_{2}-\cos (2\phi)\, (\tau_2-\tau_1)}{2(1-\tau_1)(1-\tau_2)} \, , \quad \Delta_{2}:=\frac{1-\tau_{1}\tau_{2}+\cos( 2\phi)\, (\tau_2-\tau_1)}{2(1-\tau_1)(1-\tau_2)} \, , \\
& \ell:=\frac{\sin (2\phi)\,(\tau_2-\tau_1)}{(1-\tau_1)(1-\tau_2)} \, , \quad \tau_{j}:=\frac{\xi_{j}}{\vert\xi_{j}\vert}\operatorname{tanh}\vert\xi_{j}\vert \, , \quad j\in\{1,2\} \, .
\label{nonsep-WF-para1}
\end{aligned}
\end{equation}
These depend only on the squeezing and mixing parameters $\xi_{j}$ and $\phi$, respectively. On the other hand, the coefficients proportional to the linear terms in $x_{1}, x_2$ are given by
\begin{equation}
\ell_{1}:=-2\Delta_{1}\frac{q_1}{\lambda_1}-\ell\frac{q_2}{\lambda_2}-\mathrm{i}\frac{\lambda_{1}}{\hbar}p_1 \, , \quad \ell_{2}:=\ell\frac{q_1}{\lambda_1}+2\Delta_{2}\frac{q_2}{\lambda_2}+\mathrm{i}\frac{\lambda_2}{\hbar}p_2 \, ,
\label{nonsep-WF-para2}
\end{equation}
which have an explicit dependence on the phase-space variables $q_{1}$, $q_{2}$, $p_{1}$, and $p_{2}$.

After some calculations involving elementary integrals with Gaussian functions, we explicitly determine the normalisation factor as
\begin{equation}
\mathcal{N}(\vec{q},\vec{p};\vec{\xi},\phi):=\left(\frac{\Delta}{4\pi^{2}\lambda_{1}^{2}\lambda_{2}^{2}}\right)^{\frac{1}{4}} \, 
e^{\frac{4}{\Delta}\left( \operatorname{Re}[\ell]\operatorname{Re}[\ell_{1}]\operatorname{Re}[\ell_{2}] - \operatorname{Re}[\ell_{2}]^{2}\operatorname{Re}[\Delta_{1}] -  \operatorname{Re}[\ell_{1}]^{2}\operatorname{Re}[\Delta_{2}] \right)} \, ,
\label{nonsep-WF-norm}
\end{equation}
with
\begin{multline}
\Delta:=16\operatorname{Re}[\Delta_{1}]\operatorname{Re}[\Delta_{2}]-4\operatorname{Re}[\ell]^{2})=\\
\frac{(1-\vert\tau_{1}\vert^2)(1-\vert\tau_2\vert^2)(1+\vert\tau_1\vert^2-2\operatorname{Re}[\tau_1])(1+\vert\tau_2\vert^2-2\operatorname{Re}[\tau_2])}{\vert 1-\tau_1\vert^2 \vert 1-\tau_2\vert^2} \, . 
\label{nonsep-Delta}
\end{multline}

For brevity we omit the Fock expansion as it cannot be conveniently simplified. Instead, we can use the wavefunction representation to prove that the resolution of the identity is satisfied with a uniform measure $\mu(\vec{q},\vec{p};\vec{\xi},\phi)=1$. See App.~\ref{sec:nonsep-identity} for details. We thus have
\begin{multline}
\langle\widetilde{\Psi} \vert \mathbb{I} \vert \Psi\rangle = \int_{\mathbb{R}^{4}}\mathrm{d}\vec{x}\mathrm{d}\vec{x}'[\widetilde{\Psi}(\vec{x}')]^{*}\Psi(\vec{x})
\int_{\mathbb{R}^{4}}\frac{d^{2}\vec{\alpha}}{(2\pi\hbar)^{2}} \psi(\vec{\alpha};\vec{\xi},\phi;\vec{x}')\psi(\vec{\alpha};\vec{\xi},\phi;\vec{x}) = \\
\int_{\mathbb{R}^{4}}\mathrm{d}\vec{x}'\mathrm{d}\vec{x} \, [\widetilde{\Psi}(\vec{x}')]^{*}\Psi(\vec{x}) \delta(\vec{x}-\vec{x}')=\langle\Psi'\vert\Psi \rangle \, .
\end{multline}

In this form, the quantisation and semiclassical picture of any function $f(\vec{\alpha})\equiv f(\vec{q},\vec{p})$ can be determined in the same way as the separable case because the measure in both instances is the same. That is, we have the quantisation map
\begin{equation}
f(\vec{q},\vec{p})\mapsto \hat{A}_{f}:=\int_{\mathbb{R}^{4}}\frac{\mathrm{d}\vec{q}\mathrm{d}\vec{p}}{(2\pi\hbar)^{2}}f(\vec{q},\vec{p})\vert\vec{q},\vec{p};\vec{\xi},\phi\rangle\langle\vec{q},\vec{p};\vec{\xi},\phi\vert \, ,
\end{equation}
and its alternative form through the kernel representation
\begin{equation}
\left(A^{(op)}_{f}\Psi\right)(\vec{x})=\int_{\mathbb{R}^{2}}\mathrm{d}\vec{x}\,' \mathcal{K}_{f}(\vec{\xi},\phi;\vec{x},\vec{x}\,')\Psi(\vec{x}\,') \, ,
\end{equation}
where
\begin{equation}
\mathcal{K}_{f}(\xi;\vec{x},\vec{x}\,'):=\int_{\mathbb{R}^{2}}\frac{\mathrm{d}\vec{q}\mathrm{d}\vec{p}}{2\pi\hbar}f(\vec{q},\vec{p})\psi^{*}(\vec{q},\vec{p};\vec{\xi},\phi;\vec{x}\,') \psi(\vec{q},\vec{p};\vec{\xi},\phi;\vec{x}) \, .
\label{nonsep-kernel2}
\end{equation}
In order to expose the differences between quantisations using the separable and non-separable squeezed states we consider a few examples. In Table.~\ref{Table1} we consider linear functions on the classical position $q_{1}$ and $q_{2}$, where we observe that the separable case produces a factorisable quantisation, that is, for a classical function $f(\vec{q},\vec{p})=q_{1}q_{2}$ the resulting operator is the product of the independent quadratures $\hat{x}_{1}$ and $\hat{x}_{2}$. However, the non-separable case shows that the resulting operator is not factorisable and becomes quadratic combinations of both quadratures $\hat{x}_{1}$ and $\hat{x}_{2}$. Similarly, the function $f(\vec{q},\vec{p})=q_{j}$, for $j=1,2$, the resulting operator leads to a linear combination of both quadratures as well. In the limiting case $\ell=0$ (see cases above), the resulting quantisation reduces to that of separable squeezed states.
\begin{table}
\centering
\begin{tabular}{|c|c|c|}
	\hline
	Classical & Separable SS & Non-separable SS \\
$f(\vec{q},\vec{p})$ &  $\hat{A}_{f}$ & $\hat{A}_{f}$\\
\hline
$1$ & $\mathbb{I}$ & $\mathbb{I}$ \\
\hline
$q_{1}$ & $\hat{x}_{1}$ & $ \left(1+8\frac{\operatorname{Re}[\ell]^{2}}{\Delta} \right)\hat{x}_{1} + 16 \frac{\lambda_{1}}{\lambda_{2}}\frac{\operatorname{Re}[\ell]\operatorname{Re}[\Delta_{2}]}{\Delta}\hat{x}_{2}$ \\
\hline
$q_{2}$ & $\hat{x}_{2}$ & $ \left(1+8\frac{\operatorname{Re}[\ell]^{2}}{\Delta} \right)\hat{x}_{2} + 16 \frac{\lambda_{2}}{\lambda_{1}}\frac{\operatorname{Re}[\ell]\operatorname{Re}[\Delta_{1}]}{\Delta}\hat{x}_{1}$ \\
\hline
$q_{1}q_{2}$ & $\hat{x}_{1}\hat{x}_{2}$ & $\left(1+\frac{2^{9}\operatorname{Re}[\Delta_{1}]\operatorname{Re}[\Delta_{2}]\operatorname{Re}[\ell]^2}{\Delta^{3}}\right)\hat{x}_{1}\hat{x}_{2}+\frac{4^{2}\operatorname{Re}[\Delta_{1}]\operatorname{Re}[\ell]}{\Delta}\left(1+8\frac{\operatorname{Re}[\ell]^{2}}{\Delta}\right)\frac{\lambda_{2}}{\lambda_{1}}\hat{x}_{1}^{2}+$ \\
 & & $\frac{4^{2}\operatorname{Re}[\Delta_{2}]\operatorname{Re}[\ell]}{\Delta}\left(1+8\frac{\operatorname{Re}[\ell]^{2}}{\Delta}\right)\frac{\lambda_{1}}{\lambda_{2}}\hat{x}_{2}^{2}+\frac{8\lambda_{1}\lambda_{2}\operatorname{Re}[\ell]}{\Delta^{2}}\left(3+16\frac{\operatorname{Re}[\ell]^2}{\Delta}\right)$\\
 \hline
\end{tabular}
\caption{Two-mode quantisation associated with separable and non-separable squeezed states for different classical functions $f(\vec{q},\vec{p})$.}
\label{Table1}
\end{table}

On the other hand, the semiclassical portrait is given by
\begin{equation}
f(\vec{q},\vec{p}) \mapsto \widecheck{A}_{f(\vec{q},\vec{p})}=\langle\hat{A}_{f}\rangle=\int_{\mathbb{R}^{4}}\frac{\mathrm{d}^{2}\vec{q}\mathrm{d}^{2}\vec{p}}{(2\pi\hbar)^{2}}f(\vec{q}\,',\vec{p}\,') \vert\langle \vec{q} \, ',\vec{p} \, ';\vec{\xi},\phi \vert \vec{q},\vec{p};\vec{\xi},\phi \rangle\vert^{2} \, .
\label{semiclass-nonsep}
\end{equation}
where the overlap between two non-separable squeezed states is explicitly given by
\begin{equation}
\begin{aligned}
& \vert\langle \vec{q} \, ',\vec{p} \, ';\vec{\xi},\phi \vert \vec{q},\vec{p};\vec{\xi},\phi \rangle\vert^{2} = \exp\left(\frac{\vec{R}^{T}\cdot \widetilde{\mathbb{M}} \cdot \vec{R}}{\Delta}\right) \, , \\ 
& \vec{R}:=\left( \frac{q_{1}-q_{1}'}{\lambda_{1}}, \frac{\lambda_{1}(p_{1}-p_{1}')}{\hbar},\frac{q_{2}-q_{2}'}{\lambda_{2}},\frac{\lambda_{2}(p_{2}-p_{2}')}{\hbar} \right)^{T} \, ,
\end{aligned}
\label{overlap-nonsep}
\end{equation}
with the matrix
\begin{equation}
\widetilde{\mathbb{M}}=
\begin{pmatrix}
\theta_{1} & \frac{L_{11}}{2} & \frac{\theta_{12}}{2} & \frac{L_{12}}{2}\\
\frac{L_{11}}{2} & \Xi_{1} & \frac{L_{21}}{2} & \frac{\Xi_{12}}{2}\\
\frac{\theta_{12}}{2} & \frac{L_{21}}{2} & \theta_{2} & \frac{L_{22}}{2} \\
\frac{L_{12}}{2} & \frac{\Xi_{12}}{2} & \frac{L_{22}}{2} & \Xi_{2}
\end{pmatrix}
\end{equation}
together with the coefficients
\begin{equation}
\begin{aligned}
&\theta_{1}:= 4\operatorname{Re}[\Delta_{1}]\vert\ell\vert^{2}+16\operatorname{Re}[\Delta_{2}]\vert\Delta_{1}\vert^{2}+8\operatorname{Re}[\ell]\operatorname{Re}[\Delta_{1}\ell^{*}] \, , \\
&\theta_{2}:= 16\operatorname{Re}[\Delta_{1}]\vert\Delta_{2}\vert^{2}+4\operatorname{Re}[\Delta_{2}]\vert\ell\vert^{2}+8\operatorname{Re}[\ell]\operatorname{Re}[\Delta_{2}\ell^{*}] \, , \\
&\theta_{12}:= 16\operatorname{Re}[\Delta_{1}]\operatorname{Re}[\ell\Delta_{2}]+4\operatorname{Re}[\Delta_{2}]\vert\ell\vert^{2}-8\operatorname{Re}[\ell]\operatorname{Re}[\Delta_{2}\ell^{*}] \, , \\
& \Xi_{1}:= 4\operatorname{Re}[\Delta_{2}] \, , \quad \Xi_{2}:= 4\operatorname{Re}[\Delta_{1}] \, , \quad \Xi_{12}:= 4\operatorname{Re}[\ell] \, , \\
&L_{11}:= -16\operatorname{Im}[\Delta_{1}]\operatorname{Re}[\Delta_{2}]-4\operatorname{Im}[\ell]\operatorname{Re}[\ell] \, , \\
&L_{12}:= -8\operatorname{Im}[\ell]\operatorname{Re}[\Delta_{1}]-8\operatorname{Im}[\Delta_{1}]\operatorname{Re}[\ell] \, , \\
&L_{21}:= 8\operatorname{Im}[\ell]\operatorname{Re}[\Delta_{2}]+8\operatorname{Im}[\Delta_{2}]\operatorname{Re}[\ell] \, , \\
&L_{22}:= -16\operatorname{Im}[\Delta_{2}]\operatorname{Re}[\Delta_{1}]-4\operatorname{Im}[\ell]\operatorname{Re}[\ell] \, .
\label{nonsep-coeff}
\end{aligned}
\end{equation}
Notice that for the non-separable states~\eqref{overlap-nonsep}, besides mixing the canonical positions $q_{1}$ and $q_{2}$ among themselves, they also mix the canonical position $q_{1}$ with both of the canonical momenta $p_{1}$ and $p_{2}$. The same is true vice-versa for $q_{2}$. 

From the coefficients in~\eqref{nonsep-coeff} we can identify two interesting limiting cases:

$\bullet$ If the squeezing parameters are both equal, $\tau_{1}=\tau_{2}$, we get $\ell=0$. We therefore have $\theta_{12}=\Xi_{12}=L_{12}=L_{21}=0$, and thus the overlap~\eqref{overlap-nonsep} just couples $q_{1}$ with $p_{1}$, and $q_{2}$ with $p_{2}$. That is, the semiclassical canonical position observables $q_{1}$ and $q_{2}$ couple only with their respective canonical momenta.

$\bullet$ If $\tau_{1},\tau_{2}\in\mathbb{R}$, we get $\operatorname{Im}[\Delta_{1}]=\operatorname{Im}[\Delta_{2}]=\operatorname{Im}[\ell]=0$. Therefore, the canonical position $q_{1}$ couples with $q_{2}$, and the canonical momentum $p_{1}$ couples with $p_{2}$.

$\bullet$ For $\phi=0$ and $\tau_{1},\tau_{2}\in\mathbb{R}$, the matrix $\widetilde{\mathbb{M}}$ becomes diagonal and no coupling among the semiclassical observables is generated. This corresponds to the separable squeezed state limiting case.

To illustrate these results, let us consider the classical function $f(\vec{q},\vec{p})\equiv h(\vec{q})$ which leads to 
\begin{equation}
\widecheck{A}_{h(\vec{q})}:=\frac{\pi\hbar^2}{2\sqrt{\Delta}}\int_{\mathbb{R}} \frac{\mathrm{d}\vec{q}\,'}{(2\pi\hbar)^{2}} \, h(\vec{q} \, ')e^{-\frac{\mathfrak{C}_{1}}{\lambda_{1}^{2}\Delta}(q_1-q_1')^{2} -\frac{\mathfrak{C}_{2}}{\lambda_{2}^{2}\Delta}(q_2-q_2')^{2} + \frac{\mathfrak{C}_{12}}{\lambda_{1}\lambda_{2}}(q_{1}-q_{1}')(q_{2}-q_{2}')} \, ,
\label{coupled-hq}
\end{equation}
where
\begin{equation}
\begin{aligned}
& \mathfrak{C}_{1}:=\theta_{1}-\frac{\Xi_{1}L_{12}^{2}+\Xi_{2}L_{11}^{2}+\Xi_{12}L_{11}L_{12}}{4\Delta} \, , \quad \mathfrak{C}_{2}:=\theta_{2}-\frac{\Xi_{1}L_{22}^{2}+\Xi_{2}L_{21}^{2}+\Xi_{12}L_{21}L_{22}}{4\Delta}  \\
& \mathfrak{C}_{12}:=\theta_{12}+\frac{2\Xi_{1}L_{12}L_{22}+2\Xi_{2} L_{11}L_{21}+\Xi_{12}(L_{11}L_{22}+L_{12}L_{21})}{4\Delta} \, .
\end{aligned}
\end{equation}
Notice that the Gaussian function in~\eqref{coupled-hq}, besides regularising the classical function $h(q)$, couples the canonical position $q_{1}$ with $q_{2}$. This will lead to an anisotropic semiclassical portrait $\widecheck{A}_{h(\vec{q})}$ even if the original classical function is isotropic. 

\section{Position-dependent mass models}
\label{sec:PDM}
In this section, we apply the discussion from the previous sections to a specific problem. In particular, we focus on a position-dependent mass (PDM) model defined in a classical constrained geometry. Before proceeding with our specific model, we require some generalities in both the classical and semiclassical cases. To begin with, let us consider a two-dimensional classical Hamiltonian of the form
\begin{equation}
H=H_{1}+H_{2} \, , \quad H_{j}=\frac{p_{j}^{2}}{2m_{j}(q_{j})}+V_{j}(q_{j}) \, , \quad j=1,2 \, ,
\label{H1H2}
\end{equation}
which is separable as the sum of two one-dimensional Hamiltonians. From the Hamilton equations of motion~\cite{Gol11}, we obtain the canonical momentum $p_{j}=m_{j}(q_{j})\dot{q_{j}}$, from which, the corresponding equation of motion for the position coordinate becomes
\begin{equation}
\frac{\mathrm{d}^{2}q_{j}}{\mathrm{d}t^{2}}+\frac{1}{2m_{j}(q_{j})}\left( \frac{\partial m_{j}(q_{j})}{\partial q_{j}}\right)\left( \frac{\mathrm{d}q_{j}(t)}{\mathrm{d}t}\right)^{2}+\frac{1}{m_{j}(q_{j})}\left( \frac{\partial V_{j}(q_{j})}{\partial q_{j}} \right)=0 \, , \quad j=1,2 \, .
\label{motionPDM}
\end{equation}
Note that, in the constant mass case, $m_{j}'=0$, the equations of motion~\eqref{motionPDM} reduce to the Newton equation of motion, $m_{j}\ddot{q}_{j}=-\frac{\partial V_{j}(q_{j})}{\partial q_{j}}$.

From the setup described in Sec.~\ref{sec:sep-squeezed-quantization}, the corresponding semiclassical portrait can be determined. In particular, the semiclassical Hamiltonian becomes
\begin{equation} \widecheck{H}(\vec{q},\vec{p})=\frac{1}{2}\widecheck{A}_{p_{1}^{2}[m_{1}(q_{1})]^{-1}}+\frac{1}{2}\widecheck{A}_{p_{2}^{2}[m_{2}(q_{2})]^{-1}}+\widecheck{A}_{V_{1}(q_{1})}+\widecheck{A}_{V_{2}(q_{2})} \, ,
\label{PDM-semi-H}
\end{equation}
where a general formula for $\widecheck{A}_{p_{j}^{2}h(\vec{q})}$ is given in~\eqref{sep-semi-p2h}. 

Interestingly, the semiclassical portrait admits a symplectic structure similar to that of the classical model. That is, from the semiclassical Hamiltonian~\eqref{PDM-semi-H}, we can determine the evolution of $q_{j}(t)$ and $p_{j}(t)$ through the Hamilton equations of motion (see~\cite{Gaz20a} for a detailed proof)
\begin{equation}
\dot{q}_{j}(\vec{q},\vec{p})=\frac{\partial \widecheck{H}(\vec{q},\vec{p})}{\partial p_{j}} \, , \quad 
-\dot{p}_{j}(\vec{q},\vec{p})=\frac{\partial \widecheck{H}(\vec{q},\vec{p})}{\partial q_{j}} \, , \quad j=1,2 \, ,
\label{PDM-semi-eqsmot}
\end{equation}
where $\dot{q}_{j}\equiv\frac{\mathrm{d} q_{j}}{\mathrm{d}t}$ and $\dot{p}_{j}\equiv\frac{\mathrm{d} p_{j}}{\mathrm{d}t}$. 

In the latter, time derivatives are functions of $q_{j}$ and $p_{j}$, which, as in classical Hamiltonian mechanics, may be cast into equations of motion for $q_{j}(t)$ as functions of time. To this end, we use~\eqref{sep-semi-p2h} to get
\begin{equation}
\dot{q}_{j}=p_{j}\widecheck{A}_{\mathfrak{M}_{j}} + \frac{4\hbar\gamma_{j}}{\Delta_{p_{j}}^{2}\lambda_{j}^{2}} \left( q_{j}\widecheck{A}_{\mathfrak{M}_{j}}-\widecheck{A}_{q_{j}\mathfrak{M}_{j}} \right) \, ,
\end{equation}
from which one may determine a relation between the semiclassical momentum $p_{j}$ and the velocity $\dot{q}_{j}$. To determine the equation of motion for $q_{j}$ we use the time evolution relation for any semiclassical observable $\frac{\mathrm{d} f}{\mathrm{d}t}=\{f,\widecheck{H}\}_{PB}+\frac{\partial f}{\partial t}$, with $\{f,g\}_{PB}$ the Poisson brackets. Using the latter with $\dot{q}$, and after some calculations, we get a nonlinear coupled second-order differential equation for $q_{1}$ and $q_{2}$. An explicit form will be shown in the following section. 

We have the general equations to determine the dynamics at both the classical and semiclassical levels. Their solutions are specified by the mass, potential energy, and initial conditions. One may foresee that the resulting equations of motion are in general nonlinear, and we thus have to rely on numerical calculations in most cases.

\subsection{Variable mass oscillator in constrained geometry}
\label{sec:varmass-osc}
In order to implement the results obtained so far, let us consider the PDM Hamiltonian introduced in~\cite{Gaz20a}, which is in turn contained in the family of non-linear oscillators in~\cite{Mat74}. We thus introduce the corresponding two-dimensional classical Hamiltonian
\begin{equation}
H=H_{1}+H_{2}\, , \quad H_{j}=\frac{p^{2}_{j}}{2m_{j}(q_{j})}+\overline{V}_{j}q_{j}^{2} \, , \quad m_{j}(q_{j})=\frac{m_{0}}{1-\Lambda_{j}^{2}q_{j}^{2}} \, , \quad \overline{V}_{j},\Lambda_{j}\in\mathbb{R} \, , \quad j=1,2,
\label{nl-osc1}
\end{equation}
where $m_{0}>0$ is the mass and an external oscillator interaction has been added, which can be turned off by fixing $\overline{V}_{j}=0$. Notice that the model is only well-defined inside the interval $q_{j}\in(-\Lambda_{j}^{-1},\Lambda_{j}^{-1})$, as outside of such interval the mass takes negative values. Considering the latter, we constrain the model to be defined only in the physically allowed regions. This is done by implementing a characteristic function $\chi_{E}(\vec{q})$ of the form
\begin{equation}
\chi_{E}(\vec{q}):=
\begin{cases}
1 \quad & q_{1}\in\left(-\frac{1}{\Lambda_{1}},\frac{1}{\Lambda_{1}}\right) \, , q_{2}\in\left(-\frac{1}{\Lambda_{2}},\frac{1}{\Lambda_{2}}\right) \\
0 & \textnormal{otherwise}
\end{cases}
\, ,
\label{chi-rect}
\end{equation}
so that the redefined nonlinear oscillator Hamiltonian becomes 
\begin{equation}
H_{\chi}(\vec{q},\vec{p})=\frac{\mathfrak{M}_{1}(\vec{q})p_{1}^2}{2}+\frac{\mathfrak{M}_{2}(\vec{q})p_{2}^{2}}{2}+\chi_{E}(\vec{q})\left(\overline{V}_{1}q_{1}^{2}+\overline{V}_{2}q_{2}^{2}\right) \, ,
\label{nl-osc2}
\end{equation}
with
\begin{equation}
\mathfrak{M}_{j}(\vec{q}):=\frac{\chi_{E}(\vec{q})}{m_{j}(q_{j})}=\frac{\chi_{E}(\vec{q})\left(1-\Lambda_{j}^{2}q_{j}^{2}\right)}{m_{0}} \, , \quad j=1,2.
\label{chi-mass}
\end{equation}
That is, we have introduced the characteristic function so that the dynamics are constrained to the rectangle defined by $\chi_{E}$. 

In particular, for a null oscillator interaction, $\overline{V}_{1}=\overline{V}_{2}=0$, the equations of motion~\eqref{motionPDM} for the Hamiltonian~\eqref{nl-osc1} can be determined in a closed form, leading to the solutions
\begin{equation}
q_{j}(t)=\frac{1}{\Lambda_{j}}\sin\left( \frac{\Lambda_{j} v_{0;j}t}{\sqrt{1-\Lambda_{j}^{2}q_{0;j}^{2}}}+\arcsin(\Lambda_{j} q_{0;j}) \right) \, ,
\label{q1d}
\end{equation}
with $q_{0;j}\equiv q_{j}(t=0)$ and $v_{0;j}\equiv\dot{q}_{j}(t=0)$ the initial position and velocity, respectively. 

\begin{figure}
\centering
\subfloat[][$\Lambda_{1}=1$]{\includegraphics[width=0.3\textwidth]{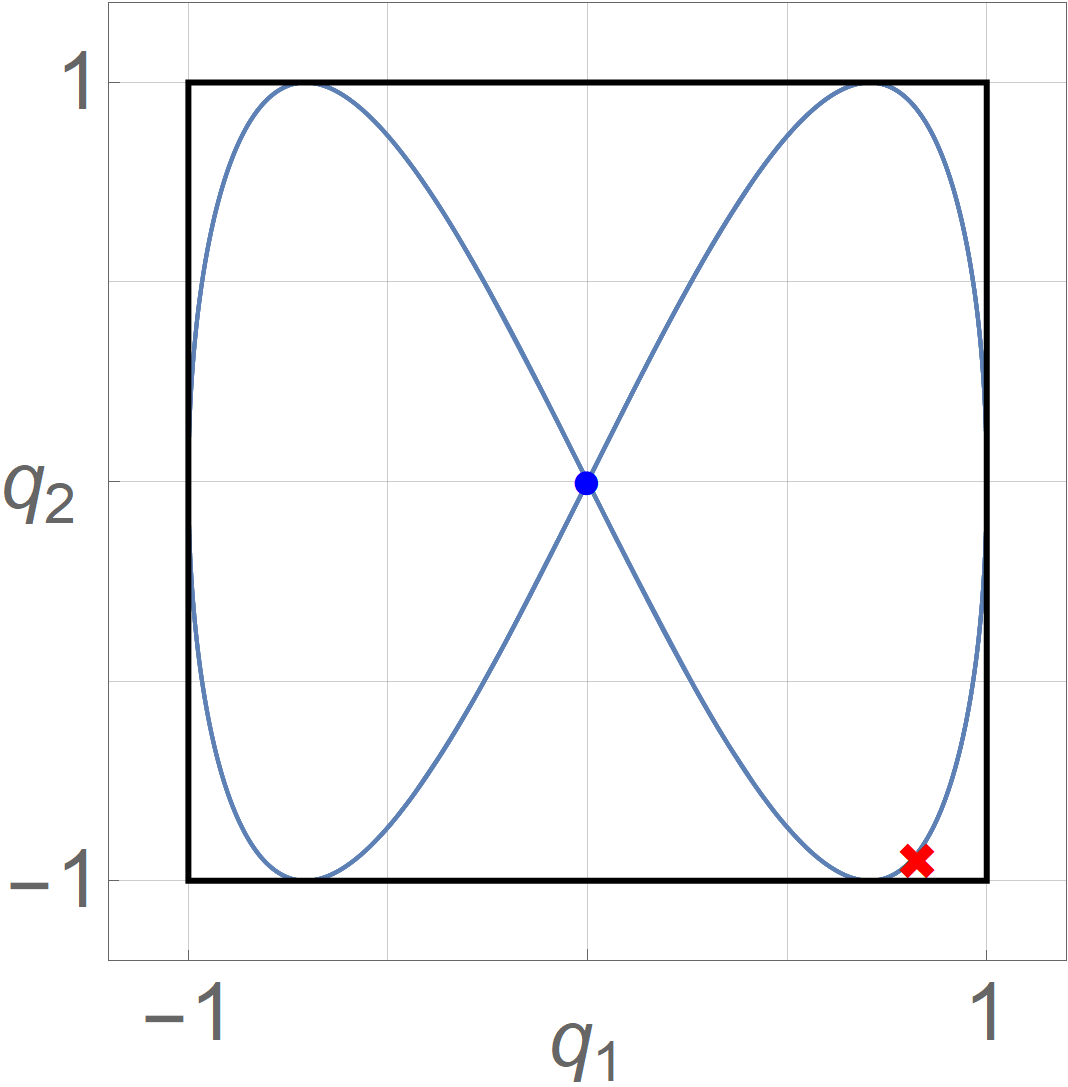}
\label{fig:varmass1-1}}
\hspace{2mm}
\subfloat[][$\Lambda_{1}=\frac{3}{2}$]{\includegraphics[width=0.3\textwidth]{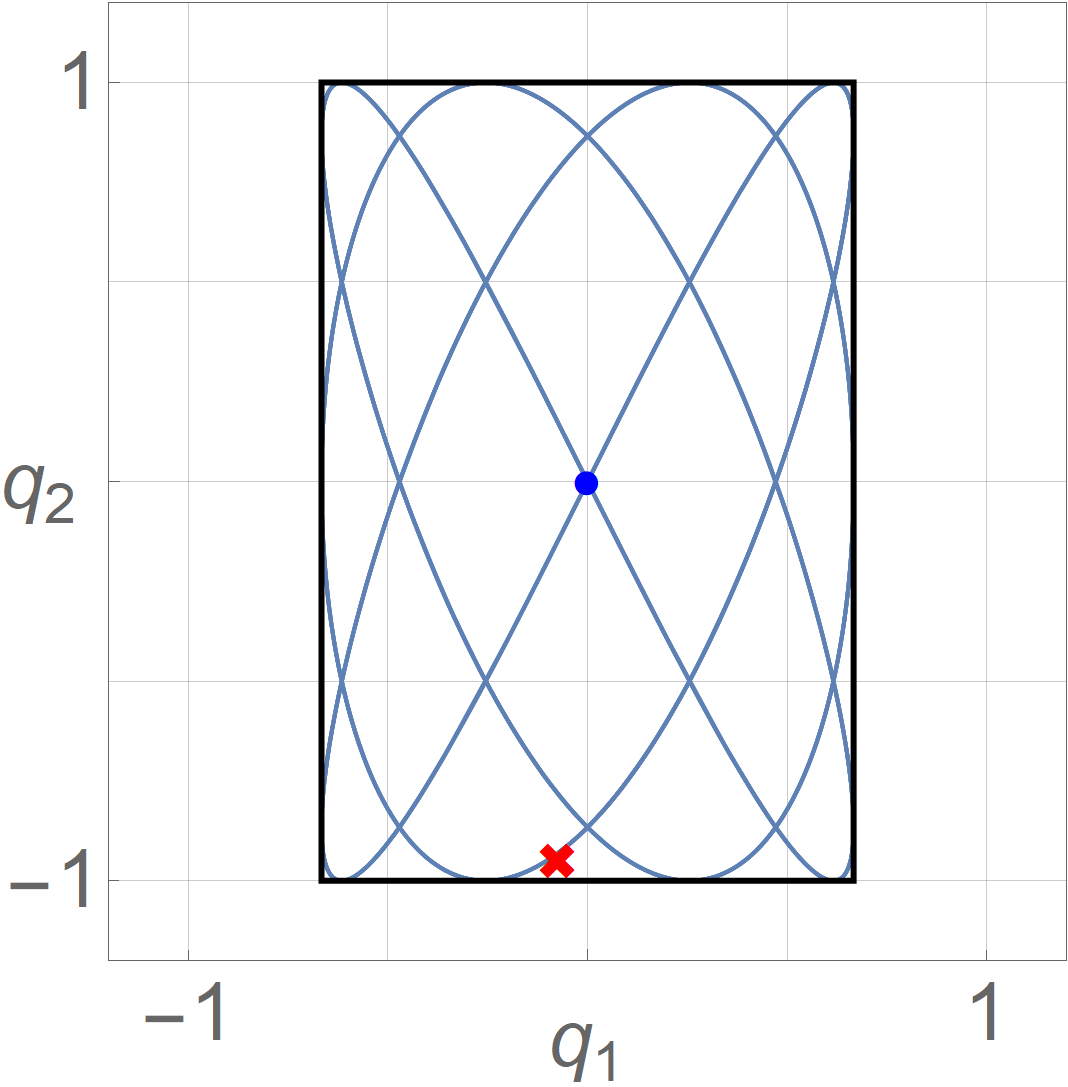}
\label{fig:varmass1-2}}
\hspace{2mm}
\subfloat[][$\Lambda_{1}=2$]{\includegraphics[width=0.3\textwidth]{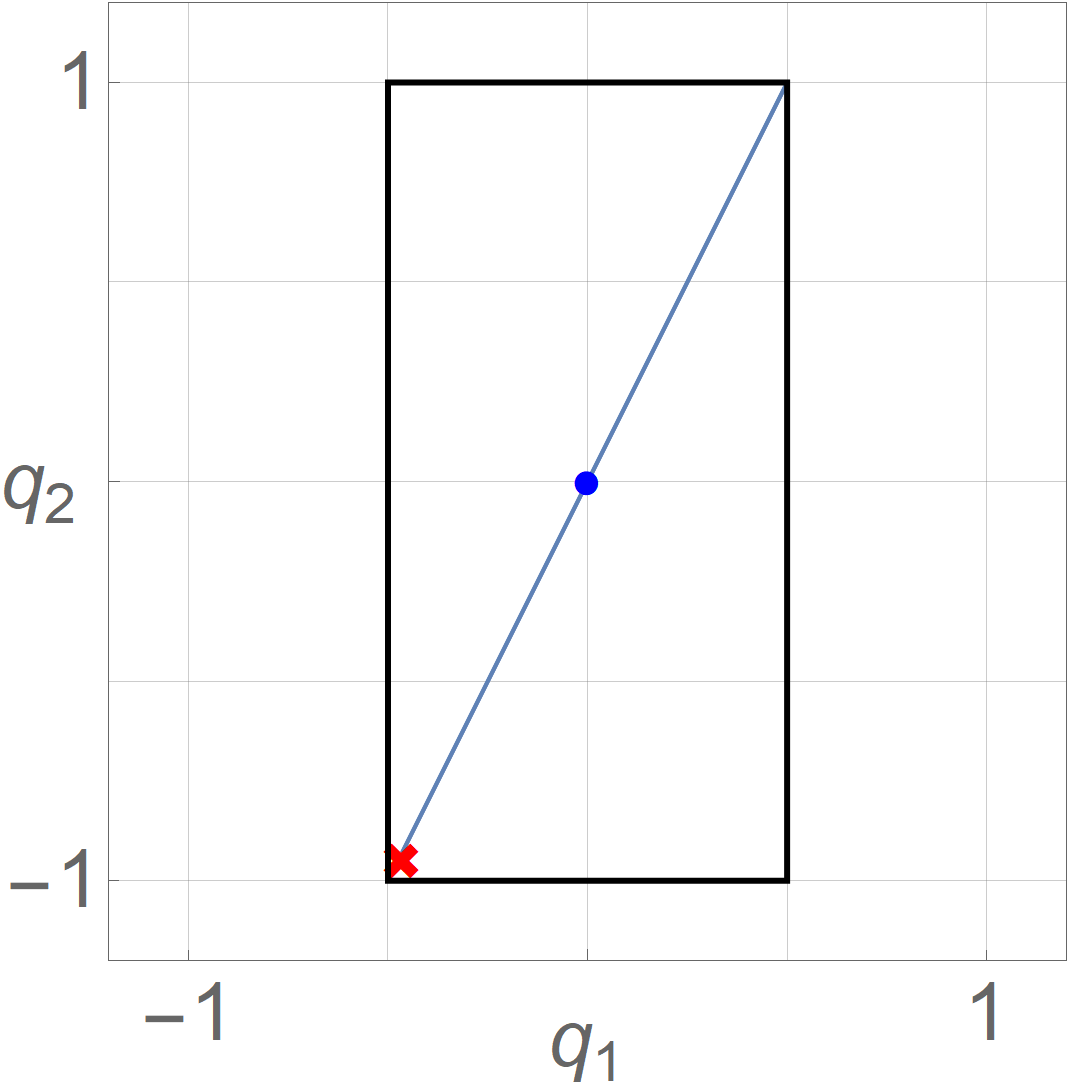}
\label{fig:varmass1-3}}
\caption{(Parameters free of units) Trajectories on the $(q_{1},q_{2})$ plane for a classical particle described by the Hamiltonian~\eqref{nl-osc2}. In every case, we have fixed $m_{0}=1$ and $\overline{V}_{1}=\overline{V}_{2}=0$ (null external oscillator interaction), together with the initial conditions $q_{0;1}=q_{0;2}=0$, $v_{0;1}=1$, and $v_{0;2}=2$. The particle is confined to the rectangle characterized by $\Lambda_{2}=1$ and the indicated values. The blue-dot and red-cross mark the initial and final positions, respectively, for the time interval $t\in(0,65)$.}
\label{fig:varmass1}
\end{figure}

Interestingly, despite the lack of a trapping interaction, the solutions for $q_{j}(t)$ describe bounded and oscillatory trajectories, which is reminiscent of the dynamics of the harmonic oscillator. The corresponding dynamics in the $q_{1}-q_{2}$ plane is depicted in Fig.~\ref{fig:varmass1} for several geometries, governed by the parameters $\Lambda_{j}$, and fixed initial conditions $q_{0;j}$ and $v_{0;j}$. In all the cases, the initial positions have been placed at the origin, $q_{0;1}=q_{0;2}=0$, whereas the ratio between the initial velocities, $\frac{v_{0;1}}{v_{0;2}}$, and the rectangle lengths, $\Lambda_{1}/\Lambda_{2}$, have been chosen so that they are both rational numbers. From the exact solution~\eqref{q1d}, it can be seen that the oscillation frequency in each direction reduces to $\Lambda_{j}v_{0;j}$, and so the ratio of the frequencies is a rational number. This explains the closed trajectories observed Figs.~\ref{fig:varmass1-1}-\ref{fig:varmass1-3}. 

\begin{figure}
	\centering
	\subfloat[][]{\includegraphics[width=0.3\textwidth]{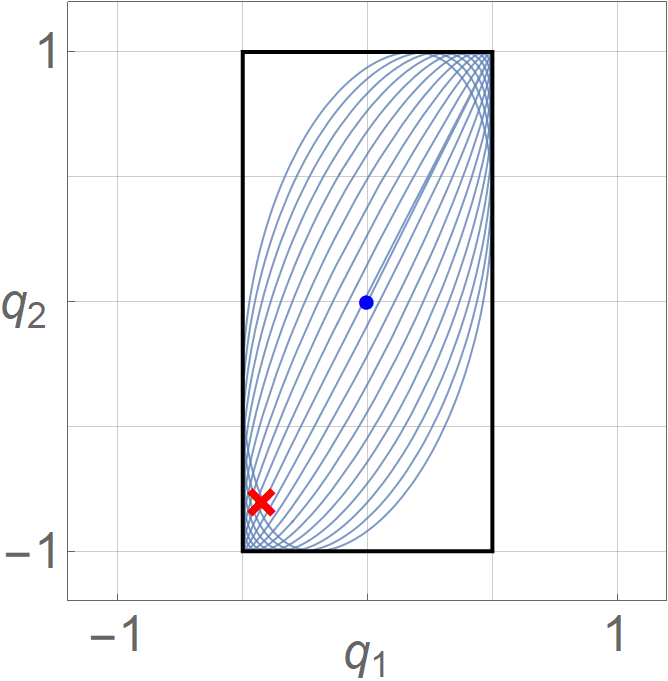}
		\label{fig:varmass1-4}}
	\hspace{2mm}
	\subfloat[][]{\includegraphics[width=0.3\textwidth]{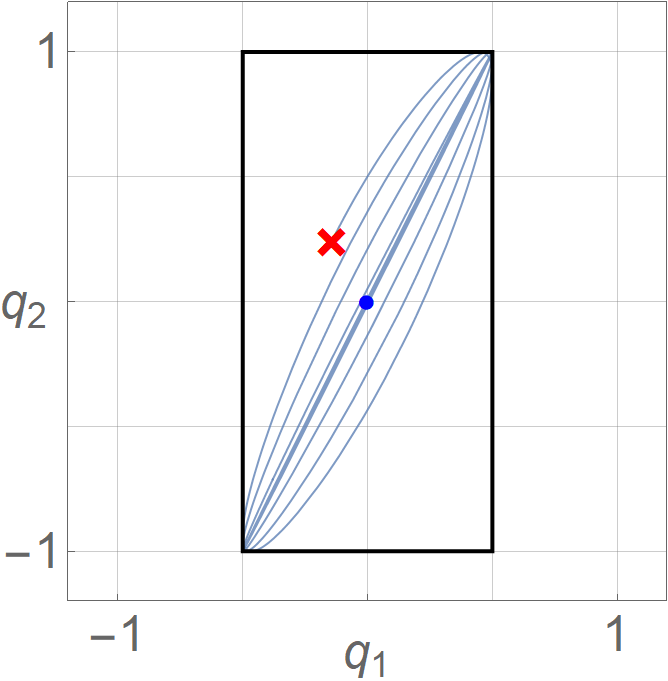}
		\label{fig:varmass1-5}}
	\hspace{2mm}
	\subfloat[][]{\includegraphics[width=0.3\textwidth]{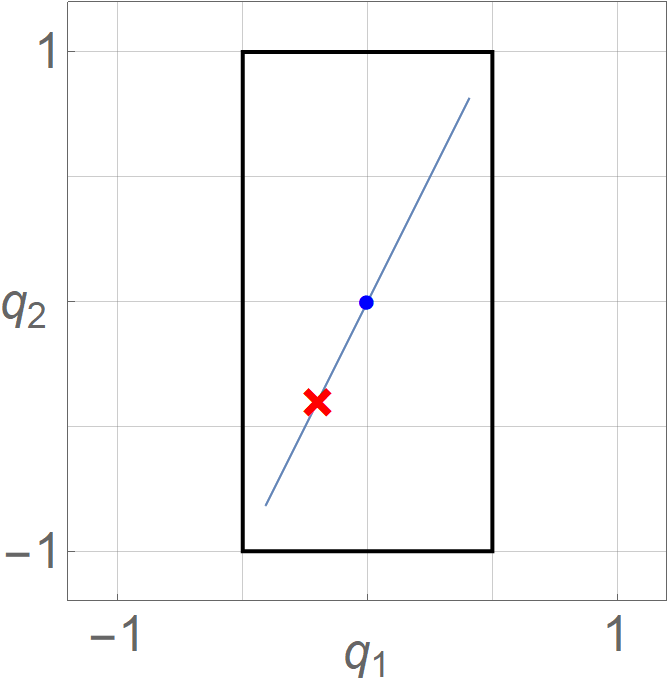}
		\label{fig:varmass1-6}}
\caption{(Parameters free of units) Classical trajectories on the $(q_{1},q_{2})$ plane for a particle described by the Hamiltonian~\eqref{nl-osc2}. In every case, we have fixed $m_{0}=5$, $\Lambda_{1}=2$, $\Lambda_{2}=1$, with the initial conditions $q_{0;1}=q_{0;2}=0$, $v_{0;1}=1$, and $v_{0;2}=2$, which corresponds to the setup in Fig.~\ref{fig:varmass1-3}. In addition, we have considered the external oscillator strength $\overline{V}_{1}=\overline{V}_{2}=1$ (a), $\overline{V}_{1}=\overline{V}_{2}=2$ (b), and $\overline{V}_{1}=\overline{V}_{2}=15$ (c). The blue-dot and red-cross represent the initial and final positions, respectively, for the time interval $t\in(0,35)$.}
	\label{fig:varmass2}
\end{figure}

On the other hand, the presence of the external oscillator interaction prevents us from obtaining a closed expression, and the dynamics cannot not be foreseen a priori. Still, it is expected that trajectories in this case would no longer be closed, as the additional nonlinearities in the equation of motion would break the strict balance required to obtain closed trajectories. Such behaviour is depicted in Fig.~\ref{fig:varmass2}, where solutions for $q_{j}(t)$ have been determined by numerical means. In these figures, we can see that the trajectories keep spreading from the null-interaction case (Fig. \ref{fig:varmass1-3}). For a large enough oscillator strength (Fig.~\ref{fig:varmass1-6}), the trajectory seems to match the one related to null-interaction case; however, the length of the trajectory shortens. This behaviour is due the additional confinement produced by the external interaction, which constrains the particle in a smaller region inside the rectangle.

\subsubsection*{Semiclassical dynamics}
To determine the semiclassical counterpart of the classical Hamiltonian~\eqref{nl-osc2}, we consider for simplicity the case $\gamma_{1}=\gamma_{2}=0$ in~\eqref{sep-semi-p2h}. In this form, we obtain the semiclassical Hamiltonian
\begin{equation}
\widecheck{H}(q_{1},q_{2};p_{1},p_{2})=\frac{p_{1}^{2}}{2}\widecheck{A}_{\mathfrak{M}_{1}}(\vec{q})+\frac{p_{2}^{2}}{2}\widecheck{A}_{\mathfrak{M}_{2}}(\vec{q})+V_{\textnormal{eff}}(\vec{q}) \, ,
\end{equation}
with the effective potential
\begin{equation} V_{\textnormal{eff}}(\vec{q}):=\frac{\hbar^{2}}{2\Delta_{p_{1}}^{2}\lambda_{1}^{2}}\widecheck{A}_{\mathfrak{M}_{1}}(\vec{q})+\frac{\hbar^{2}}{2\Delta_{p_{2}}^{2}\lambda_{2}^{2}}\widecheck{A}_{\mathfrak{M}_{2}}(\vec{q}) +\overline{V}_{1}\widecheck{A}_{q_1^{2}\chi_{E}}(\vec{q})+
\overline{V}_{2}\widecheck{A}_{q_2^{2}\chi_{E}}(\vec{q}) \, ,
\label{Veff}
\end{equation}
where the explicit form of the semiclassical portraits $\widecheck{A}_{\chi_{E}}(\vec{q})$, $\widecheck{A}_{\mathfrak{M}_{j}}(\vec{q})$, and $\widecheck{A}_{q^{2}_{j}\chi_{E}}(\vec{q})$ are presented in appendix~\ref{sec:formulae}. 

Note that the semiclassical Hamiltonian $\widecheck{H}$ has the same PDM structure as the initial classical model $H$, where the discontinuous mass and potential energy terms $\mathfrak{M}_{j}(\vec{q})$ and $V(\vec{q})$ have been replaced by their regularised counterpart $\check{A}_{\mathfrak{M}_{j}}(\vec{q})$ and $V_{\textnormal{eff}}(\vec{q})$ (see Eq.~\eqref{reg-mass}), respectively. Moreover, the effective potential term $V_{\textnormal{eff}}(\vec{q})$ is composed of the regularised truncated oscillator interaction interaction plus a term proportional to $\hbar^{2}$, which is a purely quantum correction not present in the original classical model. Interestingly, if both $\lambda_{1}>>\hbar$ and $\lambda_{2}>>\hbar$ (macroscopic scale), the purely quantum terms becomes negligible and do not play any relevant role in the dynamics. The resulting semiclassical model is still a regularised version of the original one. On the other hand, if $\lambda_{1}\sim\hbar$ and $\lambda_{2}\sim\hbar$ (small-scale), the quantum term may become dominant. In this case, one may notice that the regularised functions (see App.~\ref{sec:formulae}) are concentrated on spatial regions of the order of magnitude of $\hbar$, that is, $q_{j}\sim\hbar$ and $\Lambda_{j}^{-1}\sim\hbar$ for $j=1,2$. We focus on the small-scale regime to illustrate the purely quantum terms in the upcoming discussion.

\begin{figure}[h]
	\centering
	\subfloat[][]{\includegraphics[width=0.35\textwidth]{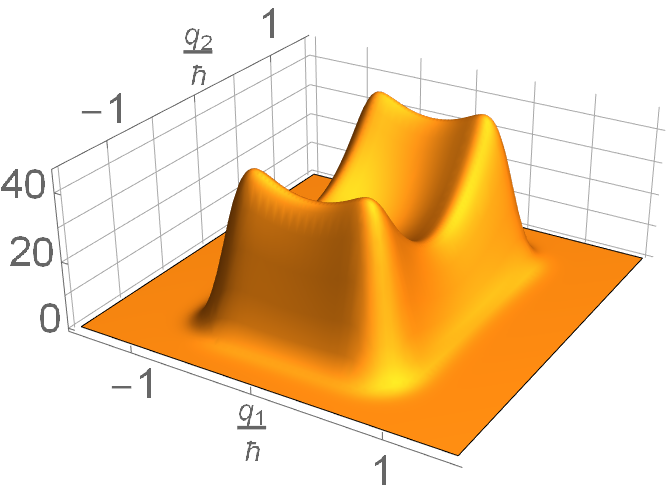}
		\label{fig:pot-density-3d}}
	\\
	\subfloat[][]{\includegraphics[width=0.25\textwidth]{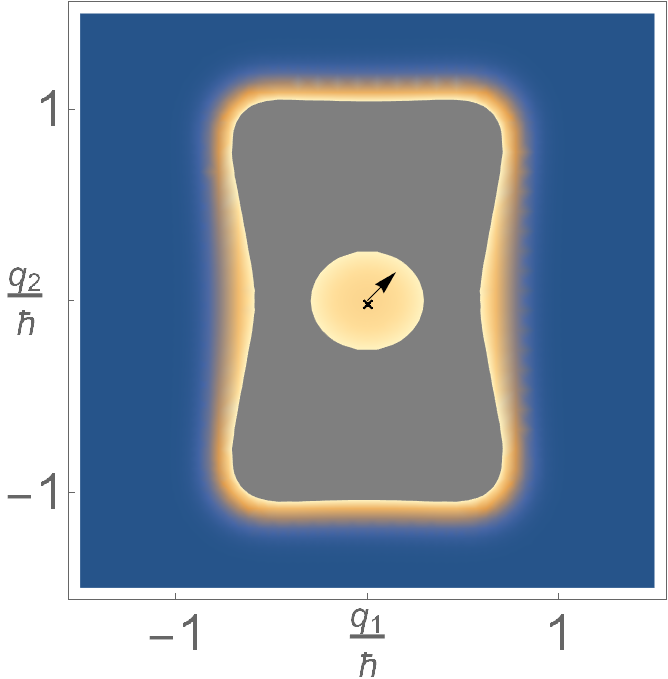}
		\label{fig:pot-density-a}}
	\subfloat[][]{\includegraphics[width=0.25\textwidth]{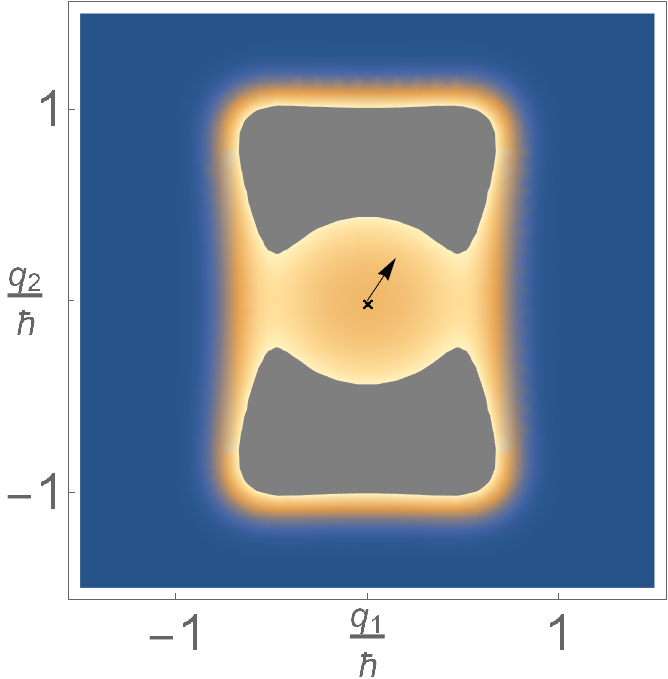}
		\label{fig:pot-density-b}}
	\subfloat[][]{\includegraphics[width=0.25\textwidth]{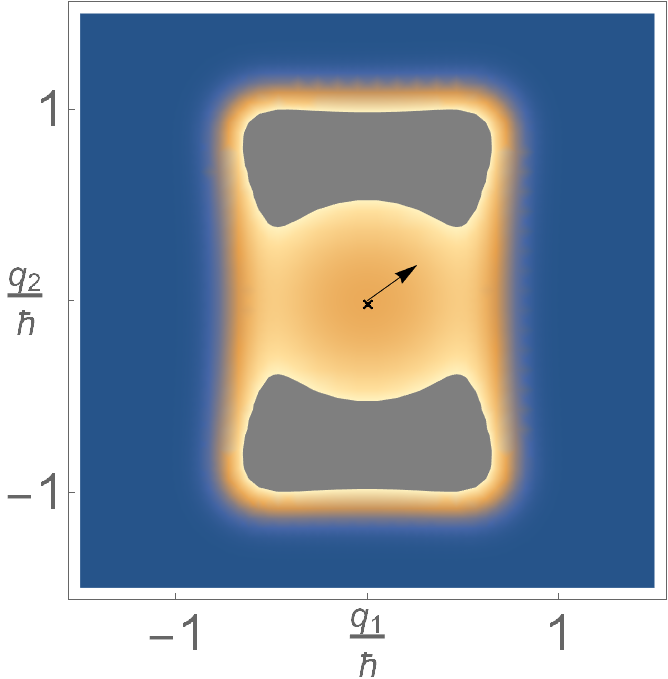}
		\label{fig:pot-density-c}}
	\caption{(Parameters free of units) (a) Regularised semiclassical effective potential~\eqref{Veff} for $m_{0}=5$, $\overline{V}_{1}=\overline{V}_{2}=50$, $\hbar\Lambda_{1}=1.5$, $\hbar\Lambda_{2}=1$, $\tau_{1}=\tau_{2}=0.9$, and $\lambda_{1}/\hbar=\lambda_{2}/\hbar=0.5$. (b)-(d) Density plot for $V_{\textnormal{eff}}(q_{1},q_{2})$, where the arrows represent the direction of the initial velocity fixed as $\{v_{0;1}/\hbar=v_{0;2}/\hbar=0.75\}$ (a), $\{v_{0;1}/\hbar=1, v_{0;2}/\hbar=1.5\}$ (b), and $\{v_{0;1}/\hbar=1.75, v_{0;2}/\hbar=1.25\}$ (c). The shadowed area denotes the interception between $V_{\textnormal{eff}}(q_{1},q_{2})$ and the energy constant $\widecheck{H}(q_{0;1},q_{0;2};p_{0;1},p_{0;2})$, with $p_{0;j}=v_{0;j}[\widecheck{A}_{\mathfrak{M}_{j}}(q_{0;1},q_{0;2})]^{-1}$ for $j=1,2$, and the aforementioned initial conditions}
	\label{fig:pot-density}
\end{figure}

Since the semiclassical Hamiltonian admits the same classical symplectic structure~\cite{Gaz20a}, one may derive the equations of motion as in the classical case. After some calculations, one arrives to
\begin{equation}
\begin{aligned}
&\ddot{q}_{1}=\frac{1}{2\widecheck{A}_{\mathfrak{M}_{1}}}\frac{\partial\widecheck{A}_{\mathfrak{M}_{1}}}{\partial q_{1}}\dot{q}_{1}^{2}-\frac{\widecheck{A}_{\mathfrak{M}_{1}}}{2\widecheck{A}_{\mathfrak{M}_{2}}^{2}}\frac{\partial\widecheck{A}_{\mathfrak{M}_{2}}}{\partial q_{1}}\dot{q}_{2}^{2}+\frac{1}{\widecheck{A}_{\mathfrak{M}_{1}}}\frac{\partial\widecheck{A}_{\mathfrak{M}_{1}}}{\partial q_{2}}\dot{q}_{1}\dot{q}_{2}-\widecheck{A}_{\mathfrak{M}_{1}}\frac{\partial V_{\textnormal{eff}}}{\partial q_{1}} \, , \\
&\ddot{q}_{2}=\frac{1}{2\widecheck{A}_{\mathfrak{M}_{2}}}\frac{\partial\widecheck{A}_{\mathfrak{M}_{1}}}{\partial q_{2}}\dot{q}_{2}^{2}-\frac{\widecheck{A}_{\mathfrak{M}_{2}}}{2\widecheck{A}_{\mathfrak{M}_{1}}^{2}}\frac{\partial\widecheck{A}_{\mathfrak{M}_{1}}}{\partial q_{2}}\dot{q}_{1}^{2}+\frac{1}{\widecheck{A}_{\mathfrak{M}_{2}}}\frac{\partial\widecheck{A}_{\mathfrak{M}_{2}}}{\partial q_{1}}\dot{q}_{1}\dot{q}_{2}-\widecheck{A}_{\mathfrak{M}_{2}}\frac{\partial V_{\textnormal{eff}}}{\partial q_{2}} \, ,
\end{aligned}
\label{PDM-semi-motion}
\end{equation}
where, for simplicity, the dependence of the semiclassical functions on $\vec{q}$ has been dropped out. Eq.~\eqref{PDM-semi-motion} defines a system of two coupled non-linear second-order differential equations and it is not possible to solve it by analytical means. Although we have to resort to numerical methods for the generation of the dynamics, we may extract some preliminary information from the effective potential. To illustrate the latter, in Fig.~\ref{fig:pot-density-3d} we plot the effective potential~\eqref{Veff} initially confined to a rectangular region defined by $\Lambda_{1}=3/2$ and $\Lambda_{2}=1$. In this case, we see that the potential barrier is much stronger in the $q_{2}$ direction compared to the barrier in the $q_{1}$ direction. Thus, we expect that at relatively low energies, particles fired towards $q_{1}$ would overcome the trapping potential and escape the confinement. This reveals that both the initial energy and direction of the particles define the dynamics, contrary to the one-dimensional case, where the energy solely dictates whether the particle gets trapped~\cite{Gaz20a}. The latter is shown in Figs.~\ref{fig:pot-density-a}-\ref{fig:pot-density-c}, where the initial velocities $v_{0;1}\equiv\dot{q}_{1}(0)$ and $v_{0;2}\equiv\dot{q}_{2}(0)$ are changed while keeping the initial positions $q_{0;1}\equiv q_{1}(0)$ and $q_{0;2}\equiv q_{2}(0)=0$ fixed. The shaded area in Figs.~\ref{fig:pot-density-a}-\ref{fig:pot-density-c} depicts the intercept between the effective potential $V_{\textnormal{eff}}(q_{1},q_{2})$ and the semiclassical energy $\widecheck{H}(q_{0;1},q_{0;2};p_{0;1},p_{0;2})$. It thus defines forbidden regions for the particle dynamics. In all cases we can relate the initial momentum and initial velocity by $p_{0;j}=v_{0;j}[\widecheck{A}_{\mathfrak{M}_{j}}(q_{0;1},q_{0;2})]^{-1}$ for $j=1,2$. In Fig.~\ref{fig:pot-density-a}, the shaded area encloses the particle, and bounded trajectories are expected. In Figs.~\ref{fig:pot-density-b}-\ref{fig:pot-density-c}, the forbidden area does not surround the particle, and both bounded and unbounded trajectories are expected depending on the direction of the initial velocity. In Fig.~\ref{fig:pot-density-b} the particle is directed towards the confining region and we expect bounded motion. However, Fig.~\ref{fig:pot-density-c} clearly shows a particle fired away from the confining region, and unbounded dynamics are expected.

\begin{figure}
\centering
\subfloat[][]{\includegraphics[width=0.3\textwidth]{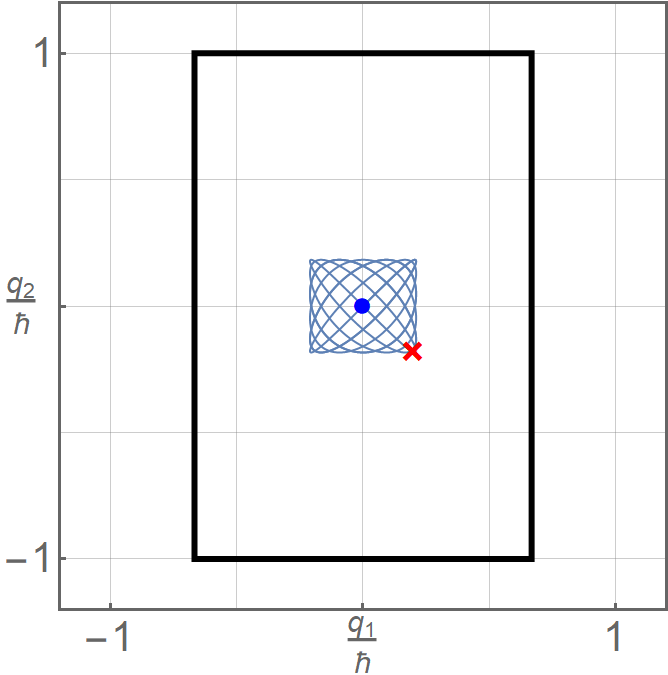}
\label{fig:avarmass1-4}}
\hspace{2mm}
\subfloat[][]{\includegraphics[width=0.3\textwidth]{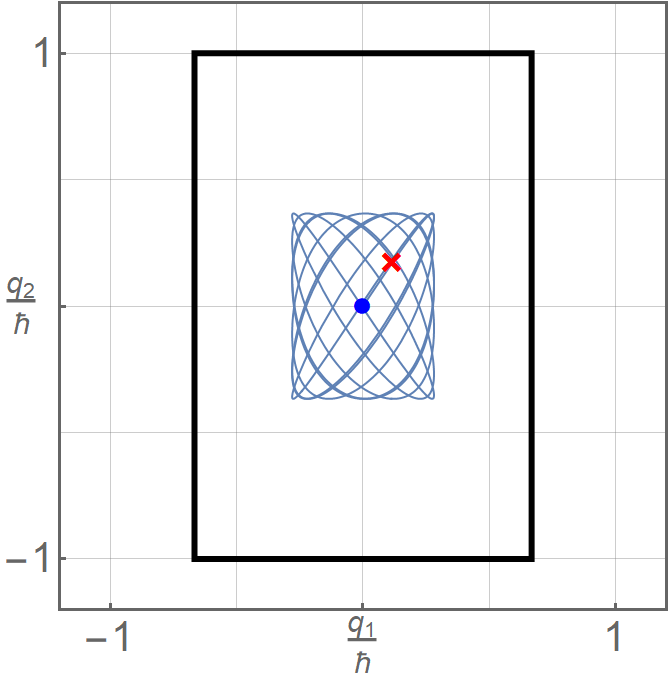}
\label{fig:avarmass1-5}}
\hspace{2mm}
\subfloat[][]{\includegraphics[width=0.3\textwidth]{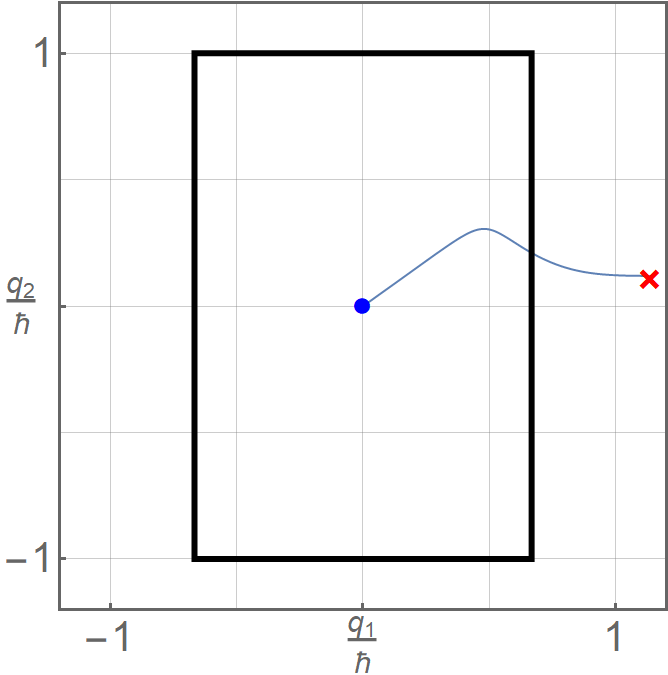}
\label{fig:avarmass1-6}}
\caption{(Parameters free of units) Trajectories obtained from the semiclassical Hamiltonian~\eqref{PDM-semi-H}. We have used $m_{0}=5$, $\hbar\Lambda_{1}=1.5$, $\hbar\Lambda_{2}=1$, $\overline{V}_{1}=\overline{V}_{2}=50$, $\lambda_{1}/\hbar=\lambda_{2}/\hbar=0.5$, $\tau_{1}=\tau_{2}=0.9$. The initial conditions are fixed to $q_{0;1}/\hbar=q_{0;2}/\hbar=0$ and $v_{0;1}/\hbar=v_{0;2}/\hbar=0.75$ (a), $v_{0;1}/\hbar=1$, $v_{0;2}/\hbar=1.5$ (a), and $v_{0;1}/\hbar=1.75$, $v_{0;2}/\hbar=1.25$ (c). The blue-dot and red-cross represent the initial and final positions, respectively, for a time interval $t\in(0,15)$.}
\label{fig:varmass3}
\end{figure}

A curious result of the regularisation is an induced force due to the wall. The regularisation of the wall has rendered it finite and the particle can escape. This is a clear departure from the classical case where the particle is always confined within the restricted region. In Figs.~\ref{fig:avarmass1-4}-\ref{fig:avarmass1-6}, we plot the dynamics corresponding to the setup in Figs.~\ref{fig:pot-density-a}-\ref{fig:pot-density-c}. In Fig.~\ref{fig:avarmass1-4} and Fig.~\ref{fig:avarmass1-5} we recover bounded motion. In the bounded case, the particle never reaches the classically confining boundary due to the smoothing of the wall. Conversely, in Fig.~\ref{fig:avarmass1-6} we see that for a particle fired away from the confining region, its initial energy and direction are enough to overcome the confining potential strength and thus it escapes.

\section{Conclusion}\label{sec.con}
In this work we have looked at quantisation in two dimensions using general families of two-dimensional squeezed states. We found the dependence of the quantised operators and their semiclassical portraits on the squeezing parameters introduced, and found that in the case where the squeezed states are coupled (entangled), an anisotropy is induced in the resulting operators and portraits. This offers additional control over the strength and direction of the regularisation of discontinuous functions when compared with standard coherent state quantisation. In principle the additional control available may allow one to quantise a theory with more precision than would be available under different quantisation schemes.

In section \ref{sec:conv-ss} we reviewed a quantisation scheme in one dimension using conventional squeezed states. In section \ref{sec:SS2D} we defined two distinct families of two-dimensional squeezed states: coordinate separable squeezed states taken as the product of two one-dimensional squeezed states, and the non-separable squeezed states which are not separable as a product of two one-dimensional squeezed states. It was found that the quantisation of the classical position functions $q_1$ and $q_2$ led to the expected quantum operators $\hat{x}_1$ and $\hat{x}_2$ respectively for the separable squeezed states. On the other hand, for the non-separable squeezed states, the quantisations of the individual classical position functions became linear combinations of the quantum position operators for both modes. In section \ref{sec:PDM} we applied the preceding formalism to a position-dependent mass model in two dimensions using the separable squeezed states, and studied a comparison between the classical and semiclassical portraits using squeezed state quantisation.

For future work, it would be interesting to look at quantisation in non-rectangular confined regions. It would seem that there are a number of physical systems one could approach when one can conveniently regularise confined regions in quantisation problems, such as the hadron bag model \cite{hadbag}, and quantum motion on non-rectangular surfaces. The anisotropic effect of the squeezing parameters would allow one to distort the restricted region to obtain semiclassical billiard-like dynamics in non-trivial geometries. Additionally, exploring quantisation in higher-dimensional systems should lead to interesting results, the classes of generalised coherent and squeezed states will proliferate and as such so will the ways in which one can quantise a problem.
\section*{Acknowledgements}
V. Hussin acknowledges the support of research grants from NSERC of Canada. J. Moran acknowledges the support of the D\'epartement de physique at the Universit\'e de Montr\'eal. K.Z. acknowledges the support from the project ``Physicists on the move II'' (KINE\'O II), Czech Republic, Grant No. CZ.02.2.69/0.0/0.0/18 053/0017163; and the funding provided by Consejo Nacional de Ciencia y Tecnolog\'ia (CONACyT), Mexico, Grant No. A1-S-24569.

\bibliographystyle{aip}
\bibliography{refsquant.bib}

\begin{thebibliography}{10}

\bibitem{gerryknight6}
C.~Gerry and P.~Knight,
\newblock {\em Introductory Quantum Optics},
\newblock Cambridge University Press, 2004.

\bibitem{Pfistera}
O.~Pfister, R.~C. Pooser, A.~S. Bradley, and M.~K. Olsen,
\newblock Frontiers in Optics {\bf OSA Technical Digest Series}, LTuF3 (2005).

\bibitem{addesso}
G.~Adesso, S.~Ragy, and A.~R. Lee,
\newblock OSID {\bf 21}, 1440001 (2014).

\bibitem{Ra2020aa}
Y.-S. Ra et~al.,
\newblock Nature Physics {\bf 16}, 144 (2020).

\bibitem{schumaker}
B.~L. Schumaker,
\newblock Physics Reports {\bf 135}, 317 (1986).

\bibitem{xinrhodes}
X.~Ma and W.~Rhodes,
\newblock Phys. Rev. A {\bf 41}, 4625 (1990).

\bibitem{glasserzhang}
W.~Zhang and R.~T. Glasser,
\newblock arXiv:2002.00323  (2020).

\bibitem{holomorphh}
S.~T. Ali, K.~Górska, A.~Horzela, and F.~H. Szafraniec,
\newblock J. Math. Phys. {\bf 55}, 012107 (2014).

\bibitem{2011}
J.~P. Gazeau and F.~H. Szafraniec,
\newblock  {\bf 44}, 495201 (2011).

\bibitem{Kla12}
J.~Klauder,
\newblock J. Phys. A: Math. Theor. {\bf 45}, 285302 (2012).

\bibitem{Pur94}
R.~Puri,
\newblock Phys. Rev. A {\bf 49}, 2178 (1994).

\bibitem{Nie97}
M.~Nieto,
\newblock Phys. Lett. A {\bf 229}, 135 (1997).

\bibitem{Mar97}
P.~Marian,
\newblock Phys. Rev. A {\bf 55}, 3051 (1997).

\bibitem{Leo11a}
R.~D. J.~L. Montel, H.~Moya-Cessa, and F.~Soto-Eguibar,
\newblock Rev. Mex. Fis. S {\bf 57}, 133 (2011).

\bibitem{Gaz21b}
J.~Gazeau, V.~Hussin, J.~Moran, and K.~Zelaya,
\newblock J. Math. Phys. {\bf 62}, 072104 (2021).

\bibitem{Moy21}
H.~Moya-Cessa and J.~Guerrero,
\newblock J. Mod. Opt. {\bf 68}, 196 (2021).

\bibitem{Alv02}
M.~N. Alvarez and V.~Hussin,
\newblock J. Math. Phys {\bf 43} (2063).

\bibitem{Zel18}
K.~Zelaya, S.~Dey, and V.~Hussin,
\newblock Phys. Lett. A {\bf 382}, 3369 (2018).

\bibitem{Zel21}
K.~Zelaya, V.~Hussin, and O.~Rosas-Ortiz,
\newblock Eur. Phys. J. Plus {\bf 136}, 534 (2021).

\bibitem{Man97}
O.~V. Man'ko and G.~Schrade,
\newblock J. Russ. Laser Res {\bf 18}, 561 (1997).

\bibitem{Ger95}
C.~C. Gerry,
\newblock J. Mod. Optics {\bf 42}, 585 (1995).

\bibitem{Thi15}
K.~Thirulogasanthar, N.~Saad, and G.~Honnouvo,
\newblock Math. Phys. Anal. Geom. {\bf 18}, 13 (2015).

\bibitem{Mor21}
J.~Moran and V.~Hussin,
\newblock Quantum Theory and Symmetries: Proceedings of the 11th International
  Symposium  (2021).

\bibitem{Gri18}
D.~J. Griffiths and D.~F. Schroeter,
\newblock {\em Introduction to Quantum Mechanics},
\newblock Cambridge University Press, 2018.

\bibitem{Zel21d}
K.~Zelaya, V.~Hussin, and O.~Rosas-Ortiz,
\newblock Eur. Phys. J. Plus {\bf 136}, 534 (2021).

\bibitem{doi:10.1119/1.16337}
J.~J. Gong and P.~K. Aravind,
\newblock American Journal of Physics {\bf 58}, 1003 (1990).

\bibitem{Eij90}
S.~J.~L. van Eijndhoven and J.~L.~H. Meyers,
\newblock J. Math. Anal. Appl {\bf 146}, 89 (1990).

\bibitem{Mou14}
M.~E.~H. Ismail and P.~Simeonov,
\newblock Proc. Am. Math. Soc {\bf 143}, 1397 (2015).

\bibitem{Gaz19}
J.~Gazeau, T.~Koide, and D.~Noguera,
\newblock J. Phys. A: Math. Theor. , 445203 (2019).

\bibitem{Gaz20a}
J.~Gazeau, V.~Hussin, J.~Moran, and K.~Zelaya,
\newblock J. Phys. A: Math. Theor. {\bf 53}, 505306 (2020).

\bibitem{Pru86}
A.~P. Prudnikov, Y.~A. Brychkov, and O.~I. Marichev,
\newblock {\em Integrals and series Volume 2: Special functions},
\newblock Gordon and Breach and Science Publisher, 1986.

\bibitem{Cre89}
P.~Crehan,
\newblock J. Phys. A: Math. Gen {\bf 22}, 811 (1989).

\bibitem{Gos16}
M.~A. de~Gosson,
\newblock {\em Born-Jordan Quantization: Theory and Applications},
\newblock Springer International Publishing, 2016.

\bibitem{Gol11}
H.~Goldstein, C.~Poole, and J.~Safko,
\newblock {\em Classical Mechanics},
\newblock Pearson Education, 2011.

\bibitem{Kim02}
M.~S. Kim, W.~Son, V.~Buzek, and P.~L. Knight,
\newblock Phys. Rev. A {\bf 65}, 032323 (2002).

\bibitem{Nie97b}
M.~Nieto and D.~Truax,
\newblock Fortschr. Phys. {\bf 45}, 145 (1997).

\bibitem{Mat74}
P.~Mathews and M.~Lakshmanan,
\newblock Q. Appl. Math. {\bf 32}, 215 (1974).

\bibitem{hadbag}
C.~E. DeTar and J.~F. Donoghue,
\newblock Annu. Rev. Nucl. Part. Sci. {\bf 33}, 235 (1983).

\end{thebibliography}



\appendix

\renewcommand{\thesection}{A}

\renewcommand{\theequation}{A-\arabic{equation}}
\setcounter{equation}{0}  

\section{Resolution of the identity through holomorphic Hermite polynomials}
\label{sec:conv-ss-holH}
In this appendix we prove the resolution of the identity associated to the one-mode squeezed states written in their Fock expansion given in~\eqref{conv-ss-expansion}. In such a case, the coefficients are written in terms of complex Hermite polynomials. Thus, to determine their completeness, we consider the holomorphic Hermite polynomials in two variables $H_{n}(x+\mathrm{i}y)$~\cite{Eij90,Mou14}, which satisfy the orthogonality relationship
\begin{equation}
\int_{\mathbb{R}^{2}}\mathrm{d}x\mathrm{d}y \, H_{n}(x+\mathrm{i}y)H_{m}(x-\mathrm{i}y)e^{-ax^{2}-by^{2}}=\frac{\pi}{\sqrt{ab}}2^{n}n!\left(\frac{a+b}{ab}\right)^{n}\delta_{n,m} \, , 
\label{HoloH-1}
\end{equation}
where the constants $a$ and $b$ are constrained by
\begin{equation}
0<a<b \, , \quad \frac{1}{a}-\frac{1}{b}=1 \, .
\label{HoloH-2}
\end{equation}

To simplify the notation, we use $u_{1}=\operatorname{Re}[\alpha]$ and $u_{2}=\operatorname{Im}[\alpha]$ throughout this section. Substituting~\eqref{conv-ss-expansion} into~\eqref{conv-ss-measure} leads to
\begin{multline}
\int_{\mathbb{R}^{2}}\frac{\mathrm{d}u_{1}\mathrm{d}u_{2}}{\pi}\mu(u_{1},u_{2})\vert \alpha(u_{1},u_{2});\xi\rangle\langle\alpha(u_{1},u_{2});\xi\vert= \\
(1-\vert\tau\vert^{2})^{\frac{1}{2}}\sum_{n,m=0}^{\infty}\frac{\tau^{\frac{n}{2}}(\tau^{*})^{\frac{m}{2}}}{(2^{n+m}n!m!)^{\frac{1}{2}}}\mathcal{F}_{n,m}\vert n\rangle\langle m\vert \, ,
\label{HoloH-ident1}
\end{multline}
where
\begin{equation}
\mathcal{F}_{n,m}=\int_{\mathbb{R}^{2}}\frac{\mathrm{d}u_{1}\mathrm{d}u_{2}}{\pi}\mu(u_{1},u_{2})e^{-(1-\operatorname{Re}[\tau])u_{1}^{2}-(1+\operatorname{Re}[\tau])u_{2}^{2}+2\operatorname{Im}[\tau]u_{1}u_{2}}
H_{n}(z_{1}+\mathrm{i}z_{2})H_{m}(z_{1}-\mathrm{i}z_{2}) \, ,
\label{HoloH-F1}
\end{equation}
and the functions $z_{1}\equiv z_{1}(u_{1},u_{2})$ and $z_{2}\equiv z_{2}(u_{1},u_{2})$ are defined through the following linear transformation:
\begin{equation}
\begin{pmatrix}
z_{1} \\
z_{2}
\end{pmatrix}
=\mathbb{M}
\begin{pmatrix}
u_{1}\\
u_{2}
\end{pmatrix} 
\, , \quad \mathbb{M}=\frac{\sqrt{1-\vert\tau\vert^{2}}}{2\vert\tau\vert}
\begin{pmatrix}
\left(\vert\tau\vert+\operatorname{Re}[\tau]\right)^{\frac{1}{2}} & \left(\vert\tau\vert-\operatorname{Re}[\tau]\right)^{\frac{1}{2}}\\
-\left(\vert\tau\vert-\operatorname{Re}[\tau]\right)^{\frac{1}{2}} & \left(\vert\tau\vert+\operatorname{Re}[\tau]\right)^{\frac{1}{2}}
\end{pmatrix} 
\, .
\end{equation}

In order to use the orthogonality~\eqref{HoloH-1}, we have make a change of variable into $z_{1}$ and $z_{2}$. In this case the differential element in the new variables is given by $\mathrm{d}u_{1}\mathrm{d}u_{2}\rightarrow\operatorname{det}(\mathbb{M}^{-1})\mathrm{d}z_{1}\mathrm{d}z_{2}$. Making these substitutions, Eq.~\eqref{HoloH-F1} becomes
\begin{equation}
\mathcal{F}_{n,m}=\frac{2\vert\tau\vert}{1-\vert\tau\vert^{2}}\int_{\mathbb{R}^{2}}\frac{\mathrm{d}z_{1}\mathrm{d}z_{2}}{\pi}\mu(z_{1},z_{2})e^{-\left(\frac{2\vert\tau\vert}{1+\vert\tau\vert}\right)z_{1}^{2}-\left(\frac{2\vert\tau\vert}{1-\vert\tau\vert}\right)z_{2}^{2}}H_{n}(z_{1}+\mathrm{i}z_{2})H_{m}(z_{1}-\mathrm{i}z_{2}) \, ,
\label{HoloH-F2}
\end{equation}
from which we realise that, in order to use the orthogonality of the holomorphic Hermite polynomials, the measure must be uniform and take the form $\mu(z_{1},z_{2})=\mu_{0}$. Moreover, from~\eqref{HoloH-F2} we identify
\begin{equation}
a\equiv\frac{2\vert\tau\vert}{1+\vert\tau\vert} \, , \quad b\equiv\frac{2\vert\tau\vert}{1-\vert\tau\vert} \, ,
\label{HoloH-ab}
\end{equation}
making it clear that the constraints in~\eqref{HoloH-2} are fulfilled for all $\vert\tau\vert<1$, or equivalently $\xi\in\mathcal{C}$. In this form, Eq.~\eqref{HoloH-1} leads to
\begin{equation}
\mathcal{F}_{n,m}=\frac{\mu_{0}}{(1-\vert\tau\vert^{2})^{\frac{1}{2}}}\left(\frac{2}{\vert\tau\vert}\right)^{n}n!\delta_{n,m} \, .
\label{HoloH-F3}
\end{equation}

Finally, by substituting~\eqref{HoloH-F3} into~\eqref{HoloH-ident1}, with $\mu_{0}=1$, we get
\begin{equation}
\int_{\mathbb{R}^{2}}\frac{\mathrm{d}u_{1}\mathrm{d}u_{2}}{\pi}\vert \alpha(u_{1},u_{2});\xi\rangle\langle\alpha(u_{1},u_{2});\xi\vert= \sum_{n=0}^{\infty}\vert n\rangle\langle n\vert=\mathbb{I} \, ,
\end{equation}
recovering the resolution of the identity for the one-mode squeezed states with respect to the uniform measure $\mu(\alpha)\equiv\mu(u_{1},u_{2})=1$.


\renewcommand{\thesection}{B}

\renewcommand{\theequation}{B-\arabic{equation}}
\setcounter{equation}{0}  

\section{Determining $\psi(\vec{q},\vec{p};\vec{\xi},\phi;\vec{x})$}
\label{sec:nonsep-WF}
In this appendix, we detail the steps followed to get the wavefunction representation associated to the two-mode states~\eqref{nonsep-SS}. The wavefunction can be determined without explicitly expanding $\vert\vec{\alpha};\vec{\xi},\phi\rangle$ in the two-mode Fock basis. We exploit the unitary transformations of the boson operators $a_1$ and $a_2$ generated by $G$ sto determine an eigenvalue equation, which in turns lead to a partial differential equation for the wavefunction. 

To begin with, we use $G$ in~\eqref{G} together with the Baker--Campbell--Hausdorff formula $e^{A}Be^{-A}=B+[A,B]+\frac{1}{2!}[A,[A,B]]+\ldots$ to get the unitary transformations
\begin{align}
& G^{\dagger}a_{1} G=\cos\phi\left(a_{1}\cosh\vert\xi_{1}\vert-a_{1}^{\dagger}\frac{\xi_{1}}{\vert\xi_{1}\vert}\sinh\vert\xi_{1}\vert \right)+\sin\phi\left(a_{2}\cosh\vert\xi_{2}\vert-a_{2}^{\dagger}\frac{\xi_{2}}{\vert\xi_{2}\vert}\sinh\vert\xi_{2}\vert \right) + \alpha_{1} \, , 
\label{appA-GDaG}\\
& G^{\dagger}a_{2} G=\cos\phi\left(a_{2}\cosh\vert\xi_{2}\vert-a_{2}^{\dagger}\frac{\xi_{2}}{\vert\xi_{2}\vert}\sinh\vert\xi_{2}\vert \right) - \sin\phi\left(a_{1}\cosh\vert\xi_{1}\vert-a_{1}^{\dagger}\frac{\xi_{1}}{\vert\xi_{1}\vert}\sinh\vert\xi_{1}\vert \right) + \alpha_{2} \, ,
\label{appA-GDbG}
\end{align}
where the transformations for the the creation operators follow straightforwardly from the latter by applying the adjoint operation and exploiting the unitarity of $G$. On the other hand, we may compute the following alternative unitary transformations:
\begin{multline}
Ga_{1}^{\dagger}G^{\dagger}=
\cos\phi\left( (a_{1}^{\dagger}-\alpha_{1}^{*})\cosh\vert\xi_{1}\vert+(a_{1}-\alpha_{1})\frac{\xi_{1}^{*}}{\vert\xi_{1}\vert}\sinh\vert\xi_{1}\vert \right)-\\
\sin\phi\left( (a_{2}^{\dagger}-\alpha_{2}^{*})\cosh\vert\xi_{1}\vert+(a_{2}-\alpha_{2})\frac{\xi_{1}^{*}}{\vert\xi_{1}\vert}\sinh\vert\xi_{1}\vert \right) \, , 
\label{appA-GaGD}
\end{multline}
\begin{multline}
Ga_{2}^{\dagger}G^{\dagger}=
\cos\phi\left( (a_{2}^{\dagger}-\alpha_{2}^{*})\cosh\vert\xi_{2}\vert+(a_{2}-\alpha_{2})\frac{\xi_{2}^{*}}{\vert\xi_{2}\vert}\sinh\vert\xi_{2}\vert \right)+\\
\sin\phi\left( (a_{1}^{\dagger}-\alpha_{1}^{*})\cosh\vert\xi_{2}\vert+(a_{1}-\alpha_{1})\frac{\xi_{2}^{*}}{\vert\xi_{2}\vert}\sinh\vert\xi_{2}\vert \right) \, .
\label{appA-GbGD}
\end{multline}

Now, by recalling that $\vert\vec{\alpha};\vec{\xi},\phi\rangle=G\vert 0,0\rangle$, we apply ~\eqref{appA-GDaG} on $\vert 0,0\rangle$, then multiply on the left by $G$ in order to get
\begin{equation}
a_{1}\vert\vec{\alpha};\vec{\xi},\phi\rangle = \left( \alpha_{1} - \frac{\xi_{1}}{\vert\xi_{1}\vert}\sinh\vert\xi_{1}\vert\cos\phi \left(Ga_{1}^{\dagger}G^{\dagger}\right)-\frac{\xi_{2}}{\vert\xi_{2}\vert}\sinh\vert\xi_{2}\vert\sin\phi \left( G a_{2}^{\dagger}G^{\dagger}\right)\right)\vert\vec{\alpha};\vec{\xi},\phi\rangle \, , 
\end{equation}
which, with the aid of~\eqref{appA-GaGD}-\eqref{appA-GbGD}, leads us to the eigenvalue equation
\begin{equation}
\left(\mathcal{A}_{1}a_{1}+\mathcal{A}_{2}a_{1}^{\dagger}+\mathcal{A}_{3}a_{2}+\mathcal{A}_{4}a_{2}^{\dagger}\right)\vert\vec{\alpha};\vec{\xi},\phi\rangle=z_{1}\vert\vec{\alpha};\vec{\xi},\phi\rangle \, ,
\label{nonsep-eigen1}
\end{equation}
where the coefficients are given by
\begin{equation}
\begin{aligned}
& \mathcal{A}_{1}:=1+\sinh^{2}\vert\xi_{1}\vert\cos^{2}\phi+\sinh^{2}\xi_{2}\sin^{2}\phi \, , \\ 
& \mathcal{A}_{2}:=\frac{\xi_{2}}{\vert\xi_{2}\vert}\sinh\vert\xi_{2}\vert\cosh\vert\xi_{2}\vert\sin^{2}\phi+\frac{\xi_{1}}{\vert\xi_{1}\vert}\sinh\vert\xi_{1}\vert\cosh\vert\xi_{1}\vert\cos^{2}\phi \, , \\
& \mathcal{A}_{3}:=\left(-\sinh^{2}\vert\xi_{1}\vert+\sinh^{2}\vert\xi_{2}\right)\sin\phi\cos\phi \, , \\
& \mathcal{A}_{4}:=\left(\frac{\xi_{2}}{\vert\xi_{2}\vert}\sinh\vert\xi_{2}\vert\cosh\vert\xi_{2}\vert-\frac{\xi_{1}}{\vert\xi_{1}\vert}\sinh\vert\xi_{1}\vert\cosh\vert\xi_{1}\vert \right)\cos\phi\sin\phi \, ,
\end{aligned}
\end{equation}
with $z_{1}=\mathcal{A}_{1}\alpha_{1}+\mathcal{A}_{2}\alpha_{1}^{*}+\mathcal{A}_{3}\alpha_{2}+\mathcal{A}_{4}\alpha_{2}^{*}$ a complex eigenvalue. Following the same steps, from the unitary transformation of $a_{2}$, we have a second eigenvalue equation of the form
\begin{equation}
\left(\mathcal{B}_{1}a_{1}+\mathcal{B}_{2}a_{1}^{\dagger}+\mathcal{B}_{3}a_{2}+\mathcal{B}_{4}a_{2}^{\dagger}\right)\vert\vec{\alpha};\vec{\xi},\phi\rangle=z_{2}\vert\vec{\alpha};\vec{\xi},\phi\rangle \, ,
\label{nonsep-eigen2}
\end{equation}
where $\mathcal{B}_{1}=\mathcal{A}_{3}^{*}$, $\mathcal{B}_{2}=\mathcal{A}_{4}$, together with
\begin{equation}
\begin{aligned}
& \mathcal{B}_{3}:=1+\sinh^{2}\vert\xi_{2}\vert\cos^{2}\phi+\sinh^{2}\xi_{1}\sin^{2}\phi \, , \\ 
& \mathcal{B}_{4}:=\frac{\xi_{2}}{\vert\xi_{2}\vert}\sinh\vert\xi_{2}\vert\cosh\vert\xi_{2}\vert\cos^{2}\phi+\frac{\xi_{1}}{\vert\xi_{1}\vert}\sinh\vert\xi_{1}\vert\cosh\vert\xi_{1}\vert\sin^{2}\phi \, .
\end{aligned}
\end{equation}
and $z_{2}=\mathcal{B}_{1}\alpha_{1}+\mathcal{B}_{2}\alpha_{1}^{*}+\mathcal{B}_{3}\alpha_{2}+\mathcal{B}_{4}\alpha_{2}^{*}$. 

From~\eqref{2d-quad}, we may revert the relationships between the boson operators and the canonical position and momentum observables such that we get 
\begin{equation}
a_{j}=\frac{\hat{x}_{j}}{\sqrt{2}\lambda_{j}}+\mathrm{i}\frac{\lambda_{j}\hat{p}_{j}}{\sqrt{2}\hbar} \, , \quad a_{j}^{\dagger}=\frac{\hat{x}_{j}}{\sqrt{2}\lambda_{j}}-\mathrm{i}\frac{\lambda_{j}\hat{p}_{j}}{\sqrt{2}\hbar} \, , \quad j=1,2 \, ,
\end{equation}
which, once substituted in both eigenvalue equation~\eqref{nonsep-eigen1} and~\eqref{nonsep-eigen2}, leads to two eigenvalue equations linear in both position $\hat{x}_{j}$ and momentum $\hat{p}_{j}$.  That is,
\begin{equation}
\begin{aligned}
\left[ \left(\frac{\mathcal{A}_{1}+\mathcal{A}_{2}}{\sqrt{2}\lambda_{1}}\right)\hat{x}_{1}+\mathrm{i}\left(\frac{\mathcal{A}_{1}-\mathcal{A}_{2}}{\sqrt{2}\hbar/\lambda_{1}}\right)\hat{p}_{1}+
\left(\frac{\mathcal{A}_{3}+\mathcal{A}_{4}}{\sqrt{2}\lambda_{2}}\right)\hat{x}_{2}+\mathrm{i}\left(\frac{\mathcal{A}_{3}-\mathcal{A}_{4}}{\sqrt{2}\hbar/\lambda_{2}}\right)\hat{p}_{2} 
-z_{1}\right]\vert\vec{\alpha};\vec{\xi},\phi\rangle=0 \, , \\
\left[ \left(\frac{\mathcal{B}_{1}+\mathcal{B}_{2}}{\sqrt{2}\lambda_{1}}\right)\hat{x}_{1}+\mathrm{i}\left(\frac{\mathcal{B}_{1}-\mathcal{B}_{2}}{\sqrt{2}\hbar/\lambda_{1}}\right)\hat{p}_{1}+
\left(\frac{\mathcal{B}_{3}+\mathcal{B}_{4}}{\sqrt{2}\lambda_{2}}\right)\hat{x}_{2}+\mathrm{i}\left(\frac{\mathcal{B}_{3}-\mathcal{B}_{4}}{\sqrt{2}\hbar/\lambda_{2}}\right)\hat{p}_{2} 
-z_{2}\right]\vert\vec{\alpha};\vec{\xi},\phi\rangle=0 \, ,
\end{aligned}
\label{nonsep-eigen3}
\end{equation}
From the latter, and using the coordinate representation $\hat{x}_{j}\equiv x_{j}$ and $\hat{p}_{j}\equiv-\mathrm{i}\hbar\partial_{x_{j}}$, we get a set of two first-order partial differential equations for $\psi(\vec{\alpha};\vec{\xi},\phi;\vec{x})$, which are solved by introducing a Gaussian ansatz of the form
\begin{equation}
\psi(\vec{\alpha};\vec{\xi},\phi;\vec{x}):=\mathcal{N}e^{-\frac{\Delta_{1}}{\lambda_{1}^{2}}x_{1}^{2}-\frac{\Delta_{2}}{\lambda_{2}^{2}}x_{2}^{2}-\frac{\ell}{\lambda_{1}\lambda_{2}}x_{1}x_{2}+\frac{\ell_{1}}{\lambda_{1}}x_{1}+\frac{\ell_{2}}{\lambda_{2}}x_{2}} \, ,
\label{nonsep-WF-ans}
\end{equation}
with $\mathcal{N}$ the normalization factor, and the unknown coefficients $\Delta_{j}$, $\ell_{j}$, and $\ell$, for $j=1,2$, are to be determined once we substitute~\eqref{nonsep-WF-ans} into both eigenvalue equations. 

This leads to a system of six equations involving the above-mentioned five unknown coefficients, which is an overdetermined system of equations. Nevertheless, the extra equation provides a compatibility condition that tells us whether the ansatz is correct. After several calculations, it can be shown that compatibility condition is fulfilled, and the ansatz~\eqref{nonsep-WF-ans} provides a valid solution. Thus, after solving the remaining five equations, we get the coefficients
\begin{equation}
\begin{aligned}
& \frac{\ell_{1}}{\sqrt{2}}=\frac{-\left( (\tau_{1}-\tau_{2})\cos 2\phi-\tau_{1}\tau_{2}+1 \right)\operatorname{Re}[\alpha_{1}] + \left( (\tau_{1}-\tau_{2})\sin 2\phi \right)\operatorname{Re}[\alpha_{2}]}{(1-\tau_{1})(1-\tau_{2})}-\mathrm{i}\operatorname{Im}[\alpha_{1}] \, , \\
& \frac{\ell_{2}}{\sqrt{2}}=\frac{-\left((\tau_{1}-\tau_{2})\sin 2\phi \right)\operatorname{Re}[\alpha_{1}] - \left((\tau_{1}-\tau_{2})\cos 2\phi+\tau_{1}\tau_{2}-1 \right)\operatorname{Re}[\alpha_{2}] }{(1-\tau_{1})(1-\tau_{2})} + \mathrm{i}\operatorname{Im}[\alpha_{2}] \, ,
\end{aligned}
\label{nonsep-WF-para3}
\end{equation}
together with $\Delta_{1}$, $\Delta_{2}$, and $\ell$ given in~\eqref{nonsep-WF-para1}. To recover the coefficients given in~\eqref{nonsep-WF-para2}, we rewrite $\alpha_{j}$ in~\eqref{nonsep-WF-para3} in terms of $q_{j}$ and $p_{j}$, with $j=1,2$, through the relationships obtained in~\eqref{sep-alpha-qp}. The normalisation constant in~\eqref{nonsep-WF-norm} follows straightforwardly by using elementary integrals involving Gaussian functions.

\renewcommand{\thesection}{C}

\renewcommand{\theequation}{C-\arabic{equation}}
\setcounter{equation}{0}  
\section{Resolution of the identity for the non-separable states}
\label{sec:nonsep-identity}
In this appendix, we explain the intermediate steps needed to recover the resolution of the identity associated with the non-separable two-mode squeezed states. The squeezed states should verify the property $\langle \Psi' \vert\mathbb{I}\vert\Psi'\rangle = \langle \Psi'\vert\Psi\rangle$, where
\begin{equation}
\mathbb{I}\equiv\int_{\mathbb{R}^{4}}\frac{\mathrm{d}\vec{q}\mathrm{d}\vec{p}}{(2\pi\hbar)^{4}} \, \mu(\vec{q},\vec{p};\vec{\xi},\phi) \, \vert\vec{q},\vec{p};\vec{\xi},\phi\rangle\langle\vec{q},\vec{p};\vec{\xi},\phi\vert \, , \quad \mathrm{d}\vec{q}=\mathrm{d}q_{1}\mathrm{d}q_{2} \, , \quad  \mathrm{d}\vec{p}=\mathrm{d}p_{1}\mathrm{d}p_{2} \, ,
\label{nonsep-ident1}
\end{equation}
to be considered an overcomplete family of states. In \eqref{nonsep-ident1}, $\mu(\vec{q},\vec{p};\vec{\xi},\phi)$ stands for the measure required to satisfy the resolution of the identity. Since the wavefunction associated to $\vert\vec{q},\vec{p};\vec{\xi},\phi\rangle$ takes the form of a two-variable non-separable Gaussian~\eqref{nonsep-WF-ans}, we consider a uniform measure $\mu(\vec{q},\vec{p};\vec{\xi},\phi)=\widetilde{\mu}(\vec{\xi},\phi)$, which accounts for any remaining constants that might appear after solving the resolution of the identity. 

Thus, using the coordinate representation in the condition $\langle \Psi'\vert\mathbb{I}\vert\Psi\rangle$ and combining with~\eqref{nonsep-ident1}, we are led to
\begin{multline}
\langle\widetilde{\Psi}\vert \mathbb{I} \vert\Psi\rangle = \int_{\mathbb{R}^{4}}\mathrm{d}\vec{x}'\mathrm{d}\vec{x} \, [\widetilde{\Psi}(\vec{x}')]^{*} \Psi(\vec{x})   \widetilde{\mu}(\vec{\xi},\phi)\left[ \left(\frac{4\operatorname{Re}[\Delta_{1}]\operatorname{Re}[\Delta_{2}-\operatorname{Re}[\ell]^{2}}{\pi^{2}} \right)^{\frac{1}{2}} \times \right. \\
e^{-\Delta_{1}x_{1}^{2}-\Delta_{1}^{*}x_{1}'^{2}-\Delta_{2}x_{2}^{2}-\Delta_{2}^{*}x_{2}'^{2}+\ell x_{1}x_{2}+\ell^{*}x_{1}'x_{2}'} \int_{\mathbb{R}^{4}}\frac{\mathrm{d}\vec{q}\mathrm{d}\vec{p}}{(2\pi\hbar)^{2}} e^{-\frac{\mathrm{i}}{\hbar}p_{1}(x_{1}-x_{1}')}e^{-\frac{\mathrm{i}}{\hbar}p_{2}(x_{2}-x_{2}')}  \\
\left. e^{-\frac{1}{\Delta}\left(\frac{\eta_{1}}{\lambda_{1}^{2}}q_{1}^{2}+\frac{\eta_{2}}{\lambda_{2}^{2}}q_{2}^{2}+\frac{\eta_{12}}{\lambda_{1}\lambda_{2}}q_{1}q_{2}\right)} e^{-\left(\frac{\ell x_{2}+\ell^{*}x_{2}'}{\lambda_{1}\lambda_{2}}+\frac{2\Delta_{1}x_{1}+2\Delta_{1}^{*}x_{1}'}{\lambda_{1}^{2}}\right)q_{1}}
e^{\left(\frac{\ell x_{1}+\ell^{*}x_{1}'}{\lambda_{1}\lambda_{2}}+\frac{2\Delta_{2}x_{2}+2\Delta_{2}^{*}x_{2}'}{\lambda_{2}^{2}}\right)q_{2}}\right] \, ,
\label{appB-2}
\end{multline}
with $\mathrm{d}\vec{x}=\mathrm{d}x_{1}\mathrm{d}x_{2}$, together with
\begin{equation}
\eta_{1}:=8\operatorname{Re}[\Delta_{1}]\left( 3\operatorname{Re}[\ell]^{2}+4\operatorname{Re}[\Delta_{1}]\operatorname{Re}[\Delta_{2}] \right) \, , \quad 
\eta_{2}:=8\operatorname{Re}[\Delta_{2}]\left( 3\operatorname{Re}[\ell]^{2}+4\operatorname{Re}[\Delta_{1}]\operatorname{Re}[\Delta_{2}] \right) \, ,
\end{equation}
and $\Delta$ given in~\eqref{nonsep-Delta}. 

From the term inside square brackets in~\eqref{appB-2}, it is clear that integrating over $p_{1}$ and $p_{2}$ leads to $2\pi\hbar \delta(x_{1}-x_{1}')$ and $2\pi\hbar \delta(x_{2}-x_{2}')$, respectively. Moreover, by integrating over $q_{1}$ and $q_{2}$ one can conclude that the term in square brackets reduces to $\delta(x_{1}-x_{1}')\delta(x_{2}-x_{2}')$. This is straightforward, as it involves elementary integrals on Gaussian functions, and will be left to the reader to verify. We thus get
\begin{equation}
\langle\widetilde{\Psi} \vert \mathbb{I} \vert\Psi\rangle = \widetilde{\mu}(\vec{\xi},\phi)\int_{\mathbb{R}^{4}}\mathrm{d}\vec{x}'\mathrm{d}\vec{x} \, [\widetilde{\Psi}(\vec{x}')]^{*}\Psi(\vec{x}) \delta(\vec{x}-\vec{x}')=\widetilde{\mu}(\vec{\xi},\phi)\langle\Psi'\vert\Psi \rangle \, ,
\end{equation}
from which it is clear that $\widetilde{\mu}(\vec{\xi},\phi)=1$ in order to fulfil the resolution of the identity. We therefore verify that the non-separable two-mode squeezed states form an overcomplete family.

\renewcommand{\thesection}{D}

\renewcommand{\theequation}{D-\arabic{equation}}
\setcounter{equation}{0}  
\section{Some useful formulae}
\label{sec:formulae}
Here we introduce the expressions for the semiclassical portraits used in Sec.~\ref{sec:varmass-osc}. We consider a characteristic function of the form
\begin{equation}
\chi_{E}(\vec{q}):=
\begin{cases}
1 \quad & q_{1}\in\left(a_{1},b_{1}\right) \, , q_{2}\in\left(a_{2},b_{2}\right) \\
0 & \textnormal{otherwise}
\end{cases}
\, .
\end{equation}
For a simplified notation, we use the reparametrised variables
\begin{equation}
z_{a_{j}}:=\frac{q_{j}-a_{j}}{\sqrt{2}\lambda_{j}\Delta_{p_{j}}} \, , \quad
z_{b_{j}}:=\frac{q_{j}-b_{j}}{\sqrt{2}\lambda_{j}\Delta_{p_{j}}} \, ,  \quad
j=1,2 \, ,
\end{equation}
so that the semiclassical portrait of the characteristic function becomes
\begin{equation}
\widecheck{A}_{\chi_{E}}(\vec{q})=\widecheck{A}_{\chi_{E_{1}}}(q_{1})\widecheck{A}_{\chi_{E_{2}}}(q_{2}) \, , \quad 
\widecheck{A}_{\chi_{E_{j}}}(q_{j}):=\frac{1}{2}\left(\operatorname{Erfc}\left( z_{b_{j}}\right) - \operatorname{Erfc}\left( z_{a_{j}}\right) \right)\, .
\label{formulae-chi}
\end{equation}
On the other hand, for the classical function $f(\vec{q},\vec{p})=q_{j}^{2}\chi_{E}(\vec{q})$ we get
\begin{equation}
\widecheck{A}_{q_{1}^{2}\chi_{E}}(\vec{q})=\widecheck{A}_{q_{1}^{2}\chi_{E_{1}}}(q_{1})\widecheck{A}_{\chi_{E_{2}}}(q_{2}) \, , \quad \widecheck{A}_{q_{2}^{2}\chi_{E}}(\vec{q})=\widecheck{A}_{q_{2}^{2}\chi_{E_{2}}}(q_{2})\widecheck{A}_{\chi_{E_{1}}}(q_{1}) \, ,
\label{formulae-power-q}
\end{equation}
with
\begin{equation}
\widecheck{A}_{q_{j}^{2}\chi_{E_{j}}}(q_{j}):=\left(q_{j}^{2}+\Delta_{p_{j}}^{2}\lambda_{j}^{2} \right)\widecheck{A}_{\chi_{E_{j}}}(q_j)+\frac{\Delta_{p_{j}}\lambda_{j}}{\sqrt{2\pi}}\left( (a_{j}+q_{j})e^{-z_{a_{j}}^{2}} - (b_{j}+q_{j}) e^{-z_{b_{j}}^{2}}\right) \, .
\end{equation}
From the latter, the semiclassical portrait of the mass functions~\eqref{chi-mass} are constructed by fixing $b_{1}=-a_{1}=\Lambda_{1}^{-1}$ and $b_{2}=-a_{2}=\Lambda_{2}^{-1}$, leading to
\begin{equation}
\widecheck{A}_{\mathfrak{M}_{1}}(\vec{q})=\frac{\widecheck{A}_{\chi_{E}}(\vec{q})-\Lambda_{1}^{2}\widecheck{A}_{q_{1}^{2}\chi_{E}}(\vec{q})}{m_{0}} \, , \quad \widecheck{A}_{\mathfrak{M}_{2}}(\vec{q})=\frac{\widecheck{A}_{\chi_{E}}(\vec{q})-\Lambda_{2}^{2}\widecheck{A}_{q_{2}^{2}\chi_{E}}(\vec{q})}{m_{0}} \, .
\label{reg-mass}
\end{equation}


\end{document}